\documentclass[12pt,letterpaper]{article}%
\usepackage{amsmath}
\usepackage{amsfonts}
\usepackage{amssymb}
\usepackage{nopageno}
\usepackage{graphicx}%
\newtheorem{theorem}{Theorem}

\newtheorem{conclusion}[theorem]{Conclusion}

\newtheorem{definition}[theorem]{Definition}

\newtheorem{remark}[theorem]{Remark}

\makeatletter
\def\@oddhead{
\vbox{
\hbox to\hsize{\oddmarkA \oddmarkB \hfill \oddmarkC}
}
}
\makeatother
\def\oddmarkA{{\bf }{}}
\def\oddmarkB{}
\def\oddmarkC{\thepage}
\begin{document}

\begin{center}
\bigskip\textbf{A Topological Theory of the Physical Vacuum}{\small \ \bigskip
}

\textbf{R. M. Kiehn}

\textit{Emeritus Professor of Physics, University of Houston}

\textit{\ Retired to Mazan, France}

\textit{\ http://www.cartan.pair.com}

\bigskip

ABSTRACT
\end{center}

\begin{quotation}
This article examines how the physical presence of field energy and
particulate matter could influence the topological properties of space time.
The theory is developed in terms of vector and matrix equations of exterior
differential forms. \ The topological features and the dynamics of such
exterior differential systems are studied with respect to processes of
continuous topological evolution. \ The theory starts from the sole postulate
that field properties of a Physical Vacuum (a continuum) can be defined in
terms of a vector space domain, of maximal rank, infinitesimal neighborhoods,
that supports a Basis Frame as a 4 x 4 matrix of C2 functions with non-zero
determinant. \ The basis vectors of such Basis Frames exhibit differential
closure. \ The particle properties of the Physical Vacuum are defined in terms
of topological defects (or compliments) of the field vector space defined by
those points where the maximal rank, or non-zero determinant, condition fails.
The topological universality of a Basis Frame over infinitesimal neighborhoods
can be refined by particular choices of a subgroup structure of the Basis
Frame, [B]. \ It is remarkable that from such a universal definition of a
Physical Vacuum, specializations permit the deduction of the field structures
of all four forces, from gravity fields to Yang Mills fields, and associate
the origin of topological charge and topological spin to the Affine torsion
coefficients of the induced Cartan Connection matrix [C] of 1-forms.
\end{quotation}

\part{\bigskip{\protect\Huge The Physical Vacuum}}

\section{Preface}

In 1993-1998, Gennady Shipov\ \cite{Shipov} presented his pioneering concept
of a "Physical Vacuum" as a modified A$_{n}$ space of "Absolute Parallelism".
\ Shipov implied that a "Physical Vacuum" could have geometrical structure
specifically related to the concept of "affine" torsion.
\ Such\ non-Riemannian structures are different from those induced by Gauss
curvature effects associated with gravitational fields, for the A$_{n}$ space
defines a domain of zero Gauss curvature. \ 

This article, however, examines how the physical presence of field energy and
particulate matter could influence the \textit{topological} properties (not
only the \textit{geometrical} properties) of space time to form a "Physical
Vacuum". \ The point of departure in this article consists of three parts:

\begin{enumerate}
\item[I] Shipov's vision, that a "Physical Vacuum" is a space of Absolute
Parallelism, is extended to include a larger set of admissible systems. \ The
larger set is based on the sole requirement that \textit{infinitesimal}
neighborhoods of a "Physical Vacuum" are elements of a vector space. \ The
additional (\textit{global}) constraint of "Absolute Parallelism" is not
utilized (necessarily). \ The sole requirement implies that the field points
$\{y^{a}\}$ of the "Physical Vacuum" as a continuum support a matrix of C2
functions, with a non-zero determinant. \ This matrix of functions is defined
as the Basis Frame, $\left[  \mathbb{B(}y^{a})\right]  $, for the "Physical
Vacuum", and represents the vector space properties of infinitesimal
neighborhoods.\ \ The Basis Frame maps vector arrays of perfect differentials
into vector arrays of 1-forms. \ The basis vectors that make up the Basis
Frame for \textit{infinitesimal} neighborhoods exhibit topological,
differential closure. \ The set of admissible Basis Frames for the vector
space of infinitesimal neighborhoods is larger than the set of basis frames
for global neighborhoods, for the infinitesimal maps need not be integrable.

\item[II] In certain domains the Frame Matrix $[\mathbb{B}]$ is singular, and
then one or more of its four eigenvalues is zero. \ These singular domains (or
objects) may be viewed as topological defects of 3 (topological) dimensions or
less embedded in the field domain of a 4 dimensional "Physical Vacuum". \ They
can be thought of as condensates or particles or field discontinuities. \ The
major theme of this article examines the field properties of the "Physical
Vacuum", which is the domain free of singularities of the type\ $\det
[\mathbb{B}]=0$. \ The Basis Frame Matrix $[\mathbb{B}]$ will be assumed to
consist of C2 functions, but only C1 differentiability is required for
deriving a linear connection that defines infinitesimal differential closure.
\ If the functions are not C2, singularities can occur in second order terms,
such as curvatures (and accelerations).
\end{enumerate}

\begin{center}%
{\includegraphics[
height=2.8063in,
width=4.3223in
]%
{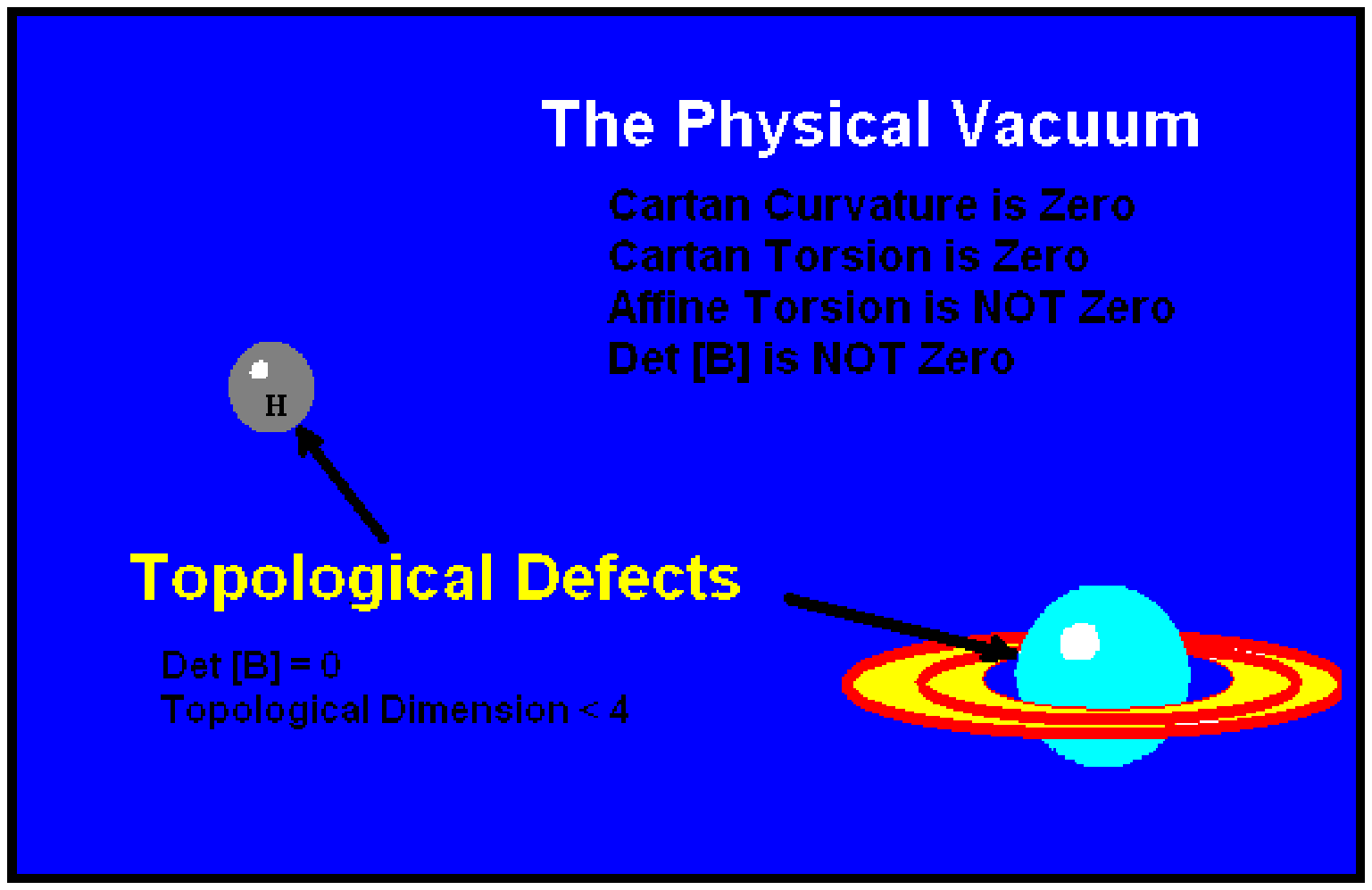}%
}%

\textbf{The 4D Physical Vacuum with 3D topological defects}
\end{center}

\begin{quote}
Although more complicated, the singular sets admit analysis, for example, in
terms of propagating discontinuities and topologically quantized period
integrals \cite{rmkperiods}. \ These topics will be considered in more detail
in a subsequent article.
\end{quote}

\begin{enumerate}
\item[III] It is recognized that topological coherent structures (fields, and
particles, along with fluctuations) in a "Physical Vacuum" can be put into
correspondence with the concepts of topological thermodynamics based on
continuous topological evolution \cite{vol1} \cite{rmkcontop}. \ Perhaps
surprising to many, topology can change continuously in terms of processes
that are not diffeomorphic. \ For example, a blob of putty can be deformed
continuously into a cylindrical rope, and then the ends can be "pasted"
together to create a non-simply connected object from a simply connected
object. \ Topological continuity requires only that the limit points of the
initial state topology be included in the closure of the topology of the final
state. \ Such continuous maps are not necessarily invertible; it is important
to remember that topology need not be conserved by such continuous processes.
\ Diffeomorphic processes require continuity of the map and its inverse and
therefor are specializations of homeomorphisms which preserve topology. \ This
observation demonstrates why tensor constraints cannot be applied to problems
of irreversible evolution and topological change \cite{rmkretro}.
\end{enumerate}

The mathematical elements used herein are expressed in terms of vector and
matrix arrays of Cartan's exterior differential forms. \ The matrix elements
of such arrays are exterior differential p-forms which are automatically
covariant with respect to diffeomorphisms. \ The combinatorial rules of
multiplication will be that of the classic row-column product associated with
the matrix multiplication of functions (0-forms). \ However, if the matrix
elements are p-forms, then the matrix "dot" product becomes the matrix
"exterior" product of differential forms, and the order of elemental factors
is preserved. \ For example:%

\begin{align}
\left[
\begin{array}
[c]{cc}%
C_{1\mu}^{1}dx^{\mu} & C_{2\mu}^{1}dx^{\mu}\\
C_{1\mu}^{2}dx^{\mu} & C_{2\mu}^{2}dx^{\mu}%
\end{array}
\right]  \symbol{94}\left\vert
\begin{array}
[c]{c}%
\sigma_{\nu}^{1}dx^{\nu}\\
\sigma_{\nu}^{2}dx^{\nu}%
\end{array}
\right\rangle  &  =\left\vert
\begin{array}
[c]{c}%
(C_{1\mu}^{1}dx^{\mu})\symbol{94}\sigma_{\nu}^{1}dx^{\nu}+(C_{2\mu}^{1}%
dx^{\mu})\symbol{94}\sigma_{\nu}^{2}dx^{\nu}\\
C_{1\mu}^{2}dx^{\mu})\symbol{94}\sigma_{\nu}^{1}dx^{\nu}+(C_{2\mu}^{2}dx^{\mu
})\symbol{94}\sigma_{\nu}^{2}dx^{\nu}%
\end{array}
\right\rangle \\
&  =\left\vert
\begin{array}
[c]{c}%
\{(C_{1\mu}^{1}\sigma_{\nu}^{1}-C_{1\nu}^{1}\sigma_{\mu}^{1})+(C_{2\mu}%
^{1}\sigma_{\nu}^{2}-C_{2\nu}^{1}\sigma_{\mu}^{2})\}dx^{\mu}\symbol{94}%
dx^{\nu}\\
\{(C_{1\mu}^{2}\sigma_{\nu}^{1}-C_{1\nu}^{2}\sigma_{\mu}^{1})+(C_{2\mu}%
^{2}\sigma_{\nu}^{2}-C_{2\nu}^{2}\sigma_{\mu}^{2})\}dx^{\mu}\symbol{94}%
dx^{\nu}%
\end{array}
\right\rangle \nonumber
\end{align}

Topological properties are inherent in the differential systems defined by
such mathematical structures. \ For example, a cohomology is defined when the
difference between two non exact p-forms is a perfect differential. \ A
classic example of cohomology is given by the first law of thermodynamics,
where the difference between the two non-exact 1-forms of Heat, $Q$, and Work,
$W$, is a perfect differential of the internal energy function:%

\begin{equation}
Q-W=dU.
\end{equation}
As another example of the topology encoded by an exterior differential system,
recall that most exact 2-forms, which satisfy the exterior differential
system, $F-dA=0,$ cannot be compact without boundary. \ Such topological
features cast doubt on the universality of "self-dual" systems.

As all p-forms can be constructed from exterior products of 1-forms, the Pfaff
topological dimension of the basis 1-forms yields pertinent topological
information. \ For example, if any 1-form is of Pfaff Topological
dimension\footnote{Thr Pfaff Topological dimension, or Class, of a 1-form
determines the minimum number of functions required to encode its topological
structure.} 3 or greater, the Cartan topology defined by such a 1-form is a
"disconnected" topology\ \cite{Baldwin}. \ For thermodynamic physical systems
that can be encoded in terms of a single 1-form of Action, the Pfaff
topological dimension can be used to distinguish between equilibrium and non
equilibrium systems and adiabatic reversible and irreversible processes
\cite{vol1}. \ Topological evolution of\ such systems can be described in
terms of Cartan's magic formulation of the Lie differential \cite{Marsden}. \ 

The topological properties of such structures and their evolutionary dynamics
can be described in terms of the Lie directional differential (which admits
topological change), but not in terms of the infamous "Covariant" differential
of tensor analysis. \ If the Lie directional differential, representing a
process, $V$, is applied to a 1-form of Action, $A$, that encodes a
thermodynamic system, the result is to produce a 1-form $Q$ of Heat.%

\begin{align}
L_{(V)}A  &  =i(V)dA+d(i(V)A)\\
&  =W+dU=Q.
\end{align}
If
\begin{equation}
L_{(V)}A=0,
\end{equation}
then the Cartan Topology is preserved. \ For integrals of n-volume forms, the
equation is related to the constraint of the variational calculus. \ \ If the
Lie derivative produces a non-zero result proportional to $nA$ then the
process is said to be homogeneous of degree $n$. \ Such a result encodes the
Sommerfeld approach to quantum mechanics, for if n is an integer:%

\begin{equation}
\text{Sommerfeld Quantum Mechanics: \ }L_{(V)}%
{\displaystyle\oint}
A=n%
{\displaystyle\oint}
A,
\end{equation}
and a quantum transition is related to an integer.

The Covariant differential is defined in terms of a "special linear
connection" which (although constructed to produce a tensor from a tensor)
does not admit irreversible topological change, a conclusion consistent with a
constraint of tensor equivalence under diffeomorphism. \ The important point
is that the covariant directional differential cannot describe non adiabatic
irreversible evolutionary processes; \ the Lie differential can. \ The Lie
differential can defined with respect to a directional field, $V^{k}, $\ where
the classic kinematic assumption fails.\
\begin{equation}
\omega^{k}\approx dx^{k}-V^{k}dt\neq0.
\end{equation}
Note that the statement\ $\omega^{k}=dx^{k}-V^{k}dt=0$ is in the form of an
exterior differential system, and imposes topological constraints on the
functions that encode the classic concept of kinematics. \ It if $\omega
^{k}\neq0$, it follows that $\omega^{k}\symbol{94}d\omega^{k}=0$, and the
Pfaff topological dimension of $\omega^{k}$ must be 2 or less. \ Processes,
$V^{k}(t),$ defined by 1 parameter group (kinematics) are always of Pfaff
Topological dimension 1, as $d\omega^{k}=dV^{k}\symbol{94}dt=0.$ In general,
however, evolutionary processes can change topological properties, such as the
Pfaff topological dimension, continuously.

In this article, be aware that the dogmatic use of the covariant differential
as an evolutionary operator is minimized in favor of the Lie differential.
\ It is useful to remark that the difference between these two concepts of
propagation by differential displacement has been described in terms of the
Higgs field \cite{atiyah}. \ It is also important to realize that the same
variety of sets can support more that 1-topological structure
\cite{rmktopstru}. \ One topological structure can be composed of different
topological substructures. \ Recall that a topological structure is
essentially enough information to define continuity. \ A process possibly
could be continuous relative to some given topological structure,\ but the
same process might not be continuous relative to a different topological
structure or substructure on the same variety of sets. \ These are not
metrical types of ideas, and do not depend upon scales, necessarily.

\section{Topological Structure of a Physical Vacuum}

\subsection{\textbf{The Fundamental Postulate:}}

Assume the existence of a matrix array of 0-forms (C2 functions), $\left[
\mathbb{B}\right]  =\left[  \mathbb{B}_{\operatorname{col}}%
^{\operatorname{row}}(y)\right]  =\left[  \mathbb{B}_{a}^{k}(y)\right]  ,$ on
a 4D variety of points $\{y^{a}\}.$ The domain for which the determinant of
$\left[  \mathbb{B}\right]  $ is not zero will be defined as a Physical Vacuum
field, and in such regions, there exists an inverse Frame, $\left[
\mathbb{B}\right]  ^{-1}.$ \ The compliment of the Physical Vacuum field
domain is defined as the singular domain, which is the realm of topologically
coherent defect structures (such as particles and field discontinuities) which
will be described later. \ From this sole topological assumption, that the
physical vacuum field is a continuous collection of infinitesimal
neighborhoods that form a vector space, the concepts displayed below are
deducible results, based on the Cartan Calculus of exterior differential
forms. \ The methods do not impose the constraints of a metric, or a linear
connection, or a coframe of 1-forms,\ as a starting point, as is done in MAG
and other theories \cite{hehl} \cite{Itin1}, but instead these concepts are
deduced in terms of topological refinements of the single hypothesis:

\begin{definition}
\textit{The Physical Vacuum is a maximal rank vector space on a 4D domain,
which, as a differential ideal, supports topological differential closure.
\ \ }
\end{definition}

\subsection{Constructive Results}

By applying algebraic and exterior differential processes developed by E.
Cartan to an element, $\left[  \mathbb{B}\right]  $, of given

equivalence class of Basis Frame of C2 functions, the following concepts are
\textbf{derived, not postulated,} results. \ It is possible to frame these
results in terms of existence theorems and constructive proofs. \ The
constructions consist of:

1. \ A "flat" Cartan Connection of 1-forms $\left[  \mathbb{C}\right]  $
leading to structural equations of curvature and torsion which are zero, but
which can support the concept of "affine" torsion. \ This derived Cartan
Connection is algebraically compatible with the Basis Frame, and is
differentially unique.

2. \ A symmetric quadratic congruence of functions, deduced by algebraic
methods from Basis Frame, that can play the role of a metric\ $\left[
g\right]  $ compatible with the Cartan Connection. \ The quadratic form can be
used to generate another connection, $\left[  \Gamma\right]  $, not equal to
the Cartan Connection, by using the Christoffel method. \ The Christoffel
connection is free of "affine" torsion, but need not be flat.

3. \ A vector array of 1-forms, $\left\vert A^{k}\right\rangle $ , which can
be put into correspondence with the classical concept of potentials used to
describe electromagnetic, hydrodynamic, and thermodynamic continua. \ The
1-forms $A^{k}$, if not zero, can have Pfaff Topological dimensions 1 to 4.
\ Exterior differentiation of $\left\vert A^{k}\right\rangle $ produces a
vector array of 2-forms, $\left\vert A^{k}\right\rangle \Rightarrow
d\left\vert A^{k}\right\rangle =\left\vert F^{k}\right\rangle ,$ that can be
put into correspondence with gauge invariant field intensities (think E and
B). \ This vector $\left\vert F^{k}\right\rangle $\ of 2-forms is "pair" and
is an invariant of differential mappings that need not be diffeomorphisms.

4. \ Another vector array of 2-forms $\left\vert G^{k}\right\rangle $ can be
deduced by exterior differentiation of the formula that describes
infinitesimal mappings of differentials $d\{\left[  \mathbb{B}\right]
\circ\left\vert dx^{k}\right\rangle \}=d\left\vert A^{k}\right\rangle
=\left\vert F^{k}\right\rangle .$ \ This vector array\ $\left\vert
G^{k}\right\rangle =\left[  \mathbb{C}\right]  \circ\left\vert dx^{k}%
\right\rangle $ is an "impair" structure\ that defines coefficients of \ field
excitations (think D and H). \ The vector array $\left\vert G^{k}\right\rangle
$, unlike the vector array of 2-forms, $\left\vert F^{k}\right\rangle $, is
not an invariant of differential mappings that are not diffeomorphisms. \ \ It
is remarkable that the coefficients of $\left\vert G^{k}\right\rangle $ are
those that have been described (historically) as the coefficients of "Affine
Torsion". \ It is further remarkable that the theory associates the Affine
Torsion of the Cartan Connection with the Field intensities of
electromagnetism. \ Without Affine Torsion, relative to a Cartan Connection,
there is no D and H excitations, no charge current densities, and no Field
intensities, E and B. \ It has been conjectured that Affine Torsion is related
to Spin, but the theory of the physical vacuum goes much deeper.

\begin{remark}
\ Affine torsion relative to a Cartan Connection for a Physical Vacuum is the
source of topological charge and topological spin.
\end{remark}

5. \ By algebraic decomposition, the Cartan Connection $\left[  \mathbb{C}%
\right]  $ can be uniquely split into a Christoffel part\ $\left[
\Gamma\right]  $\ and a Residue part $\left[  \mathbb{T}\right]  $. \ The
Affine Torsion antisymmetry is entirely contained in the Residue part. \ As
the equations of structure built on the Cartan Connection vanish, a constraint
is placed upon the contributions to the exterior differential curvature
equations of each component of the decomposition. \ The result can be
interpreted as a global equivalence principle relating to curvatures induced
by a Christoffel metric connection and the curvature 2-forms of interaction
between the Christoffel components and the Residue components of the Cartan Connection.

6. \ 3-forms of Topological Torsion, $A\symbol{94}F$ and Topological Spin,
$A\symbol{94}G$\ , can be constructed leading to quantized period integrals,
independent from macro or microscopic scales. \ The concept of the Lorentz
force is a derivable expression created by use of the Lie differential, not
the covariant differential, to describe continuous topological evolution.

7. \ The Bianchi identities are demonstrated to be statements defining the
cohomological structure of certain matrices of 3 forms, similar to the
cohomological structure of thermodynamics in terms of 1-forms.

8. \ The idea of a Higgs function is demonstrated to be related to internal
energy of a thermodynamic system.

\section{Differential Closure}

\subsection{Cartan Connections for infinitesimal neighborhoods}

\subsubsection{The Derivation of the Cartan Right Connection}

From the identity $\left[  \mathbb{B}\right]  \circ\left[  \mathbb{B}\right]
^{-1}=\left[  \mathbb{I}\right]  ,$ use exterior differentiation to
\textit{derive} the (right) Cartan Connection $\left[  \mathbb{C}\right]  $ as
a matrix of 1-forms. \ \
\begin{align}
\text{Right Cartan Connection }\text{: }  &  \left[  \mathbb{C}\right]  \text{
}\nonumber\\
d\left[  \mathbb{B}\right]  \circ\left[  \mathbb{B}\right]  ^{-1}+\left[
\mathbb{B}\right]  \circ d\left[  \mathbb{B}\right]  ^{-1}  &  =d\left[
\mathbb{I}\right]  =0\\
\text{ \ hence }d\left[  \mathbb{B}\right]   &  =\left[  \mathbb{B}\right]
\circ\left[  \mathbb{C}\right]  ,\\
\text{\ where\ }\left[  \mathbb{C}\right]   &  =-d\left[  \mathbb{B}\right]
^{-1}\circ\left[  \mathbb{B}\right] \\
&  =+\left[  \mathbb{B}\right]  ^{-1}\circ d\left[  \mathbb{B}\right] \\
&  =[C_{am}^{b}(y)dy^{m}]
\end{align}
As each of the$\ a$ columns of the Basis Frame are linearly independent, they
may be considered as Basis Vectors, $\mathbf{e}_{(a)}^{k},$ for a vector space
at the point $\{y\}.$ \ In terms of the Cartan Connection, differential of any
specific basis vector $\mathbf{e}_{(a)}^{k}$ is a linear combination of the
basis vectors $\mathbf{e}_{(b)}^{k}$\ of the set, thereby exhibiting
differential closure:
\begin{align}
&  :\text{Differential Closure}\nonumber\\
d\mathbf{e}_{(a)}^{k}  &  =\Sigma_{b}\mathbf{e}_{(b)}^{k}\{\Sigma_{m}%
C_{am}^{b}dy^{m}\}.
\end{align}
The Cartan Connection is a matrix of 1-forms; but the Basis Frame is a matrix
of functions (0-forms), not a matrix of 1-forms. \ \ The literature is a bit
confusing on this point, for often the tensor analysis machinery is used to
describe a "tetrad" by the symbols $\mathbf{\varepsilon}_{b}^{k}$ in some
places, and discuss its determinant properties as if it was an array of
functions. \ Then, in other places, the same symbol is used to describe a
matrix "coframe" of 1-forms. \ The matrix display used herein gets rid of this
confusion; \ to repeat, the Basis Frame is a matrix of functions, not 1-forms.
\ The Cartan Connection $\left[  \mathbb{C}\right]  $ is a matrix array of
1-forms. \ 

\subsubsection{Infinitesimal Neighborhoods}

The fundamental assumption of differential closure implies that the Basis
Frame $\left[  \mathbb{B}(y)\right]  $ can be used to map vector arrays of
exact differentials $\left\vert dy^{a}\right\rangle $ into infinitesimally
nearby vector arrays of 1-forms $\left\vert \sigma^{k}\right\rangle $ which
are not necessarily exact differentials $\left\vert dx^{k}\right\rangle :$%
\begin{align}
\text{A set of Infinitesimal Basis Frames }  &  :\left[  \mathbb{B}%
_{\operatorname{col}}^{\operatorname{row}}(y)\right] \\
\text{Maps 1-forms \ }\left[  \mathbb{B}_{a}^{k}(y)\right]  \circ\left\vert
dy^{a}\right\rangle \  &  \Rightarrow\ \left\vert \sigma^{k}\right\rangle .
\label{inmap}%
\end{align}
It is important to realize that the set of such Basis Frames, $\left[
\mathbb{B}(y)\right]  ,$ over infinitesimal neighborhoods is larger than the
set of "global" basis frames $\left[  \mathbb{F}(y)\right]  $ which linearly
map finite vectors into finite vectors,
\begin{align}
\text{A set of Finite Basis Frames }  &  :\left[  \mathbb{F}(y)\right]
\nonumber\\
\text{Maps 0-forms \ }\left[  \mathbb{F}(y)\right]  \circ\left\vert
y^{a}\right\rangle \  &  \Rightarrow\ \left\vert x^{k}\right\rangle .
\end{align}
The set of Basis Frames, $\left[  \mathbb{B}(y)\right]  $, does not generate
necessarily a set of diffeomorphisms.

Also note that $\left\vert \sigma^{k}\right\rangle $ is a vector array of
1-forms, and is not the same as a "coframe" matrix of 1-forms which makes up a
cornerstone of Metric-Affine-Gravity theories \cite{Itin1} \cite{hehl}. \ The
vector array of 1-forms $\left\vert \sigma^{k}\right\rangle $ can be closed,
exact, integrable, or non integrable. \ These are topological properties of
the Pfaff Topological dimension (Class) of each of the 1-forms, $\sigma^{k}.$
\
\begin{align}
Exact  &  :\sigma^{k}=dy^{k},\ \ \ \ \ \ \text{Pfaff dimension 1}\\
d\sigma^{k}  &  =d(dy^{k})=0,\ \\%
{\displaystyle\oint}
\sigma^{k}  &  =0\\
Closed~but\ not~exact  &  :d\sigma^{k}=0,\text{ \ Homogeneous degree 0}\\%
{\displaystyle\oint}
\sigma^{k}  &  =2\pi\neq0\text{ \ A Period Integral}\\
\text{ example}  &  \text{:}\sigma^{k}=\left\vert xdy-ydx\right\rangle
/\left(  x^{2}+y^{2}\right) \\
Integrable~Not~Closed  &  :d\sigma^{k}\neq0,\ \sigma^{k}\symbol{94}d\sigma
^{k}=0\ \ \ \ \ \ \ \ \ \ \text{Pfaff dimension 2}\\
Not~integrable  &  :\sigma^{k}\symbol{94}d\sigma^{k}\neq0,\ d\sigma
^{k}\symbol{94}d\sigma^{k}=0\ \ \text{Pfaff dimension 3}\\
Irreversible  &  :d\sigma^{k}\symbol{94}d\sigma^{k}\neq
0\ \ \ \ \ \ \ \ \ \ \ \ \ \ \ \ \ \ \ \ \ \ \ \ \text{Pfaff dimension 4}%
\end{align}
An important topological fact is that when the Pfaff topological dimension of
any 1-form is 3 or greater, the Cartan topology associated with such a system
is a disconnected topology \cite{vol1}. \ Moreover such systems of Pfaff
Topological dimension greater than 2 have the thermodynamic properties of non
equilibrium systems. \ Examples of these different topological possibilities
are given Part 2.

\subsubsection{The Cartan Left Connection}

It is also possible to define a left Cartan Connection as a matrix of 1-forms,
$\left[  \mathbf{\Delta}\right]  $,%
\begin{align}
\text{Left}  &  :\text{ Cartan Connection }\left[  \mathbf{\Delta}\right]
\text{\ \ }\\
d\left[  \mathbb{B}\right]   &  =\left[  \mathbf{\Delta}\right]  \circ\left[
\mathbb{B}\right]  ,\\
\left[  \mathbf{\Delta}\right]   &  =-\left[  \mathbb{B}\right]  \circ
d\left[  \mathbb{B}\right]  ^{-1}\\
&  =+d\left[  \mathbb{B}\right]  \circ\left[  \mathbb{B}\right]  ^{-1}.
\end{align}
The coefficients that make up the matrix of 1-forms, $\left[  \mathbf{\Delta
}\right]  ,$ can be associated with what have been called the Weitzenboch
connection coefficients..

The Right and Left Cartan connections are not (usually) identical. \ They are
equivalent in terms of the similarity transformation: \
\begin{equation}
\left[  \mathbb{C}\right]  =\left[  \mathbb{B}\right]  ^{-1}\circ\left[
\mathbf{\Delta}\right]  \circ\left[  \mathbb{B}\right]  .
\end{equation}
The left Cartan Connection, in general, is \textit{not} the same as the
transpose of the right Cartan Connection. \ 

Also note that inverse matrix also enjoys differential closure properties.
\begin{align}
d\left[  \mathbb{B}\right]  ^{-1}  &  =\left[  \mathbb{B}\right]  ^{-1}%
\circ\left[  -\mathbf{\Delta}\right]  ,\\
&  =\left[  -\mathbb{C}\right]  \circ\left[  \mathbb{B}\right]  ^{-1}.
\end{align}

\subsection{Electromagnetic 2-forms from a topological structure}

\subsubsection{Vectors of Intensity 2-forms \ $\left\vert F^{k}\right\rangle
$}

It is remarkable that from the sole assumption that defines the physical
vacuum, exterior differentiation leads to the topological formalism of
electromagnetism. \ Both field structures of field intensities (analogous to E
and B) and the field structures of the field excitations (analogous to D and
H) are deduced by successive applications of the exterior derivative. Both the
Maxwell Faraday equations and the Maxwell Ampere equations appear as necessary
consequences of the definition of the "Physical Vacuum".

In fact, if the 1-forms $\left\vert \sigma^{k}\right\rangle $ were written to
include a constant factor of physical dimensions, $\hslash/e$, the resulting k
individual 1-forms are formally equivalent to a set of (up to 4)
electromagnetic\ 1-forms of Action (the vector potentials), $\left\vert
A^{k}\right\rangle $. \ Suppose for k = 1 to 3, $d\sigma^{k}=dA^{k}%
\Rightarrow0,$ then the single remaining 1-form can represent (formally) the
classical development of electromagnetic field, $F=dA=d\sigma^{4}$
\cite{vol4}. \ This equation defines topological features of an exterior
differential system. \ An example is given in Section 3, below, where the
Basis Frame is presumed to be developed from the Hopf map, which has the
properties described above:
\begin{equation}
dA^{1}=dA^{2}=dA^{3}=0;\ \ dA^{4}=F.
\end{equation}

The fact that, in general, there can be 4 such electromagnetic-like structures
is a new development deduced from the single assumption used to define the
"Physical Vacuum". \ These multiple field structures under suitable
constraints of topological refinement can lead to Yang Mills theory, and
perhaps more importantly to its generalizations.

For purposes of more rapid comprehension - based on the assumption of
familiarity with electromagnetic theory - the notation for the Basis Frame
infinitesimal mapping formula is rewritten as:
\begin{equation}
\left[  \mathbb{B}_{a}^{k}(y)\right]  \circ\left\vert dy_{k}\right\rangle
\ \Rightarrow\ \left\vert A^{k}(y,dy)\right\rangle . \label{diffmap}%
\end{equation}
The notation is specific, but the formalism is universal and is valid for any
continuum structures, such as found in fluids and plasmas. \ 

Exterior differentiation of the infinitesimal mapping equation
eq(\ref{diffmap}) generates a vector of (exact) 2-forms $\left\vert
F^{k}\right\rangle $ which is formally equivalent (for each index, $k$)\ to
the (pair, or even) 2-form of $\mathbf{E,B}$ field intensities of
electromagnetic theory.%
\begin{align}
\text{Vector }\left\vert F^{k}\right\rangle  &  :\text{ of intensity 2-forms
}\nonumber\\
\left[  \mathbb{B}(y)\right]  \circ\left[  \mathbb{C}\right]  \symbol{94}%
\left\vert dy^{a}\right\rangle \  &  \Rightarrow\ \left\vert dA^{k}%
\right\rangle =\left\vert F^{k}\right\rangle ,\\
\text{Field Intensity 2-forms }  &  :\left\vert F^{k}(y,dy)\right\rangle
=\left\vert dA^{k}\right\rangle ,
\end{align}
The field intensity 2-forms (think E and B coefficients) are true tensor
invariants relative to diffeomorphisms. \
\begin{align}
\text{By diffeomorphic substitution }  &  :\text{of }\vartheta^{b}%
\Rightarrow\text{ }y^{k}=\left\vert f^{k}(\vartheta^{b}\right\rangle )\,,\\
\text{and}\left\vert dy^{k}\right\rangle  &  =\left[  \mathbb{J}\right]
\circ\left\vert d\vartheta^{k}\right\rangle ,\text{ into eq(\ref{diffmap})}\\
\text{ producing }\left[  \mathbb{B}(\vartheta)\right]   &  \Leftarrow\left[
\mathbb{B}(y)\right]  \circ\left[  \mathbb{J}\right] \\
\text{ with }\ \ \left[  \mathbb{B}(\vartheta)\right]  \circ\left\vert
d\vartheta\right\rangle \ \  &  \Leftarrow\ \left[  \mathbb{B}(y)\right]
\circ\left\vert dy\ \right\rangle \\
\text{yields }\left\vert A^{k}(\vartheta,d\vartheta)\right\rangle  &
=\left\vert A^{k}(y,dy)\right\rangle ,\\
\text{ and }\left\vert F^{k}(\vartheta,d\vartheta)\right\rangle  &
=\left\vert F^{k}(y,dy)\right\rangle .
\end{align}

\subsubsection{Vectors of Excitation 2-forms: \ $\left\vert G^{b}\right\rangle
$}

On the other hand, another vector array of two forms $\left\vert
G^{b}\right\rangle $ (think D and H coefficients) can be constructed from the formalism:%

\begin{align}
\text{Vector }\left\vert G^{b}\right\rangle  &  :\text{ of excitation 2-forms
}\nonumber\\
\left[  \mathbb{B}(y)\right]  \circ\left[  \mathbb{C}\right]  \symbol{94}%
\left\vert dy^{a}\right\rangle \  &  =\ \left[  \mathbb{B}(y)\right]
\circ\left\vert G^{b}\right\rangle ,\\
\text{Field Excitation 2-forms }  &  =\left\vert G^{b}\right\rangle =\left[
\mathbb{C}_{am}^{b}dy^{m}\right]  \symbol{94}\left\vert dy^{a}\right\rangle \\
&  =\left\vert 2\mathbb{C}_{[am]}^{b}dy^{m}\symbol{94}dy^{a}\right\rangle .
\end{align}
The "vector array" $\left\vert G^{b}\right\rangle $ of excitation 2-forms,
unlike the vector array of intensity 2-forms, $\left\vert F^{k}\right\rangle
,$ is not a\ true "tensor" invariant with respect to diffeomorphisms. \ It
transforms under pullback in terms of the inverse Jacobian mapping.%

\begin{align}
\text{By diffeomorphic }  &  :\text{substitution of }y^{a}=\left\vert
f^{a}(\vartheta^{k}\right\rangle )\,,\\
\ \ \text{\ with }dy^{a}  &  =\left[  J_{acobian}\right]  \circ\left\vert
d\vartheta^{k}\right\rangle \ \ \ \\
\ \left[  \mathbb{B}(\vartheta)\right]  \circ\left\vert G^{b}(\vartheta
,d\vartheta)\right\rangle ,  &  \Leftarrow\ \left[  \mathbb{B}(y)\right]
\circ\left\vert G^{k}(y,dy)\right\rangle ,\\
\text{yields }\left\vert G^{b}(\vartheta,d\vartheta)\right\rangle  &  =\left[
J_{acobian}\right]  ^{-1}\circ\left\vert G^{k}(y,dy)\right\rangle
\end{align}

\begin{conclusion}
Counter to current tensor dogma, the bottom line is that, in general, the
vector array of field excitations $\left\vert G^{b}\right\rangle $ is an array
of "impair" 2-forms, and depends upon a choice of diffeomorphic coordinates.
\ The vector array of field intensities $\left\vert F^{k}\right\rangle $ is an
array of "pair" 2-forms, and does not upon the choice of diffeomorphic
coordinates. \ This fact implies that charge is a pseudo-scalar, not a scalar
\cite{PostQuRe}.
\end{conclusion}

\subsection{Affine Torsion is the source of Excitation 2-forms (EM Sources)}

The vector of Excitation Torsion 2-forms $\left\vert G^{b}\right\rangle
=\left[  \mathbb{C}_{a}^{b}\right]  \symbol{94}\left\vert dy^{a}\right\rangle
\ $has coefficients that are in 1-1 correspondence with what has been defined
historically as the coefficients of "Affine" torsion. It is important to
realize that the use of the words "Affine torsion" to describe the
antisymmetric coefficients of a Cartan connection is unfortunate, and has
nothing to do with whether or not the Basis Frame matrix is a member of the
Affine group, or one of its subgroups. \ Classically, the affine group is a
\textit{transitive} group of 13 parameters in 4D, (see p.162 in Turnbull
\cite{Turnbull}). \ The anti-symmetry concept related to torsion is described
by the same formula that defines the vector of excitation 2-forms, $\left\vert
G^{b}\right\rangle =\left[  \mathbb{C}_{a}^{b}\right]  \symbol{94}\left\vert
dy^{a}\right\rangle ,$ for any Basis Frame $\left[  \mathbb{B}(y)\right]  $.
\ \ For example, the torsion\ formula holds equally well for Basis Frames
which are elements of the 15 parameter projective group, which is not affine.
\begin{align}
\text{{\small (Affine) \ }Torsion 2-forms\ \ \ }\left\vert G^{b}\right\rangle
&  =\left[  \mathbb{C}\right]  \symbol{94}\left\vert dy^{a}\right\rangle ,\\
\text{Vector of Field Excitation 2-forms \ }  &  =\left\vert G^{b}%
\right\rangle \\
&  =\left[  \mathbb{B}\right]  ^{-1}\circ\left\vert F^{k}\right\rangle ,
\end{align}
The vector of 2-forms $\left\vert G^{b}\right\rangle $ is formally equivalent
(for each index $b$)\ to the (impair, or odd) 2-form (density) of the field
excitations (D and H) in electromagnetic theory \cite{vol4}. \ In the notation
of electromagnetism, the source of field excitations (and, consequently,
topological charge and and topological spin) is due to the Affine Torsion
components of the Cartan Connection. \ 

Note that the matrix $\left[  \mathbb{B}\right]  ^{-1}$ plays the role of the
Constitutive map between $\mathbf{E,B}$ and $\mathbf{D,H}$. \
\begin{align}
\left\vert G^{b}\right\rangle  &  =\left[  \mathbb{B}\right]  ^{-1}%
\circ\left\vert F^{k}\right\rangle ,\label{CM}\\
\left[  \mathbb{B}\right]  ^{-1}  &  \approx\text{a Constitutive map}%
\end{align}
If the global (integrability) assumption, \ $\left[  \mathbb{F}(y)\right]
\circ\left\vert y^{a}\right\rangle \ \Rightarrow\ \left\vert x^{k}%
\right\rangle ,$ is imposed, then it is possible by exterior differentiation
to show that a constraint must be established between the excitation 2-forms
and the Cartan Curvature 2-forms constructed from the globally integrable
Basis Frames, $\left[  \mathbb{F}(y)\right]  :$%

\begin{align}
\left[  \mathbb{F}(y)\right]  \circ\{[\mathbb{C}_{\mathbb{F}}]\circ\left\vert
y^{a}\right\rangle \ +\left\vert y^{a}\right\rangle  &  =\ \left\vert
dx^{k}\right\rangle ,\\
\text{ such that \ }[\mathbb{C}_{\mathbb{F}}]\circ\left\vert dy^{a}%
\right\rangle  &  =-\{d[\mathbb{C}_{\mathbb{F}}]+[\mathbb{C}_{\mathbb{F}%
}]\symbol{94}[\mathbb{C}_{\mathbb{F}}]\}\circ\left\vert y^{a}\right\rangle \ .
\end{align}
Hence as the vector of excitation 2-forms $\left\vert G^{b}\right\rangle $ has
been defined in terms of the Cartan Connection, two different results are
obtained for the two different types of Basis Frames:
\begin{align}
\left\vert G^{b}\right\rangle _{\mathbb{B}}  &  =[\mathbb{C}_{\mathbb{B}%
}]\symbol{94}\left\vert dy^{a}\right\rangle \\
\left\vert G^{b}\right\rangle _{\mathbb{F}}  &  =[\mathbb{C}_{\mathbb{F}%
}]\symbol{94}\left\vert dy^{a}\right\rangle .
\end{align}
The integrability condition places a constraint on the Cartan Curvature and
the Affine torsion coefficients of the Cartan Connection, $[\mathbb{C}%
_{\mathbb{F}}]$, which is not equivalent to the constitutive map (eq \ref{CM})
given above:%

\begin{align}
\left\vert G^{b}\right\rangle _{\mathbb{F}}  &  =-\{d[\mathbb{C}%
]+[\mathbb{C}]\symbol{94}[\mathbb{C}]\}\left\vert y^{a}\right\rangle \\
&  \neq\left[  \mathbb{B}\right]  ^{-1}\circ\left\vert F^{k}\right\rangle .
\end{align}
The result demonstrates that the set of infinitesimal Basis Frames is much
different from the global set of Basis Frames.

None of this development depends upon the explicit specification of a metric,
reinforcing the fact that Maxwell's theory of Electrodynamics is a
topological, not a geometric theory \cite{vol4}. \ Again, remember that the
electromagnetic notation is used as a learning crutch to emphasize the
universal ideas of the Physical Vacuum. \ The formulas are valid topological
descriptions of the field structures of all continuum "fluids". \ 

Herein, for simplicity, it is assumed that all functions of the Basis Frame
are at least C2. \ However, note that the definitions of the matrix of
\ connection 1-forms $[\mathbb{C}]$ and the vector of 2-forms $\left\vert
F\right\rangle $ only require C1 functions.

\subsubsection{Cartan Torsion $\neq$ Affine Torsion}

It is possible to use the left matrix of Cartan Connection 1-forms to define
another vector of torsion 2-forms. \ Exterior differentiation of the
infinitesimal mapping equation eq. (\ref{diffmap}) produces the equation:
\begin{equation}
\left[  \mathbf{\Delta}_{m}^{k}\right]  \symbol{94}(\left[  \mathbb{B}_{a}%
^{m}(y)\right]  \circ\left\vert dy^{a}\right\rangle )\Rightarrow\left[
\mathbf{\Delta}_{m}^{k}\right]  \symbol{94}\left\vert \sigma^{m}\right\rangle
=\ \left\vert d\sigma^{k}\right\rangle .
\end{equation}
The algebraic combination given below, is defined as the Cartan vector of
Torsion 2-forms, $\left\vert \Sigma^{k}\right\rangle $. \ It is not the same
as the vector of excitation ("affine") Torsion 2-forms defined by $\left\vert
G^{b}\right\rangle _{\mathbb{B}}=[\mathbb{C}_{\mathbb{B}}]\symbol{94}%
\left\vert dy^{a}\right\rangle $
\begin{align}
\text{Cartan }  &  :\text{vector of Torsion 2-forms }\left\vert \Sigma
^{k}\right\rangle \\
\left\vert \Sigma^{k}\right\rangle  &  =\left\vert d\sigma^{k}\right\rangle
-\left[  \mathbf{\Delta}_{m}^{k}\right]  \symbol{94}\left\vert \sigma
^{m}\right\rangle \neq\left\vert G^{k}\right\rangle ,\label{tors2f}\\
\left\vert \Sigma^{k}\right\rangle  &  \Rightarrow0.
\end{align}
The Cartan vector of Torsion 2-forms, $\left\vert \Sigma^{k}\right\rangle $,
is always Zero for "Physical Vacuums", while the vector of excitation Torsion
2-forms\footnote{Whose coefficients are the classical "affine" torsion
coefficients.}, $\left\vert G^{k}\right\rangle $ is not necessarily zero.
\ The $\left\vert \Sigma^{k}\right\rangle $ are often said to define one of
Cartan's equations of structure. \ Another equation of structure, based upon
curvature, will be deduced below. \ Cartan's equations of structure are equal
to zero for the Physical Vacuum.

\subsubsection{Maxwell Faraday and Maxwell Ampere Equations}

The exterior derivative of the vector of field intensity 2-forms $\left\vert
F^{k}\right\rangle $ vanishes, resulting in $k$ sets of partial differential
equations of the Maxwell Faraday \cite{vol4} type: \
\begin{align}
\left\vert dF^{k}\right\rangle  &  \Rightarrow0\\
&  =\text{Maxwell Faraday PDE,s.}%
\end{align}
The concept requires the functions that define the potentials $\left\vert
A^{k}\right\rangle $ be C2 differentiable.

The exterior derivative of the vector of 2-form excitations $\left\vert
G^{b}\right\rangle $ produces a vector of 3-forms (formally) equivalent (for
each index $b$)\ to the conserved electromagnetic charge-current density for
C2 functions.%
\begin{align}
\left\vert dG^{b}\right\rangle  &  =\left\vert J^{b}\right\rangle ,\\
d\left\vert J^{b}\right\rangle  &  =0,\\
&  =\text{Maxwell Ampere PDE,s}%
\end{align}

It is extraordinary that the format of both the Maxwell Faraday and the
Maxwell Ampere equations occur naturally from exterior differential processes
applied to the fundamental postulate of a Physical Vacuum. \ If all but one of
the four 1-form components of $\left\vert A^{k}\right\rangle $ are closed,
then the formalism encodes the classical theory of Electromagnetic fields. \ A
non trivial example constructed from the Hopf map will be given in Part 2.
\ Moreover, if the totality of the four 1-form components of $\left\vert
A^{k}\right\rangle $ are not closed, the same starting point encodes the
fields that are utilized by Yang Mills theory. \ Each of these specializations
is a topological refinement.

\subsection{The Lorentz Force and the Lie differential}

The Lorentz force is a \textit{derived, universal,} concept in terms of the
thermodynamic cohomology. \ It is generated by application of Cartan's Magic
formula \cite{Marsden} to the 1-form $A$ that that encodes all or part of a
thermodynamic system. \ The system, $A$, can be interpreted as an
electromagnetic system, or a hydrodynamic system, or any other system that
supports continuous topological evolution. \cite{vol1}. \ For the purposes
herein apply Cartan's magic formula to the formula for infinitesimal mapping
produced by the matrix multiplication of a vector of perfect (exact)
differentials of the base variables given by eq(\ref{diffmap}).%

\begin{align}
\left[  \mathbb{B}_{a}^{k}(y)\right]  \circ\left\vert dy_{k}\right\rangle \
&  \Rightarrow\ \left\vert A^{k}(y,dy)\right\rangle \\
L_{(V)}\{\left[  \mathbb{B}_{a}^{k}(y)\right]  \circ\left\vert dy_{k}%
\right\rangle \}  &  =L_{(V)}\left\vert A^{k}(y,dy)\right\rangle .
\end{align}

Recall that the exterior derivative of any specific 1-form, $A^{k}$, if not
zero, can be defined as a 2-form with coefficients of the type%

\begin{align}
F  &  =dA=\{\partial A_{k}/\partial x^{j}-\partial A_{j}/\partial
x^{k}\}dx^{j}\symbol{94}dx^{k}=F_{jk}dx^{j}\symbol{94}dx^{k}\\
&  =\mathbf{B}_{z}dx\symbol{94}dy+\mathbf{B}_{x}dy\symbol{94}dz+\mathbf{B}%
_{y}dz\symbol{94}dx+\mathbf{E}_{x}dx\symbol{94}dt+\mathbf{E}_{y}%
dy\symbol{94}dt+\mathbf{E}_{z}dz\symbol{94}dt.\nonumber
\end{align}
The specialized notation for the coefficients used above is that\ often used
in studies of electromagnetism, but the topological 2-form concepts are
universal, independent from the notation. \ 

Given any process that can be expressed in terms of a vector direction field,
$V=\rho\lbrack\mathbf{V},1]$, and for a physical system, or component of a
physical system, that can be encoded in terms of a 1-form of Action, $A, $ the
topological evolution of the 1-form relative to the direction field can be
described in terms the Lie differential:%
\begin{align}
L_{(V)}A  &  =i(V)dA+d(i(V)A)\\
&  =i(V)F+d(i(V)A)\\
&  =\rho\{\mathbf{E}+\mathbf{V}\times\mathbf{B)}_{k}dx^{k}-\rho\{\mathbf{E}%
\cdot\mathbf{V}\}dt\\
&  +d(\rho\mathbf{A\cdot V}-\rho\phi)\\
&  =\text{Work due to Lorentz force - dissipative power }\\
&  \text{+ change of internal interaction energy}.
\end{align}
Note that if the notation is changed (such that the vector potential is
designated as the velocity components of a fluid), then the "Lorentz force"
represents the classic expression to be found in the\ formulation of the
hydrodynamic Lagrange Euler equations of a fluid \cite{vol3}. \ A fluid, based
upon a 1-form of Action of Pfaff Topological dimension 2 (or greater) obeys
the topological equivalent of a Maxwell Faraday induction law!

\begin{remark}
The\ universal concept of a Lorentz force is derived from the properties of a
"Physical Vacuum", and does not require a separate postulate of existence.
\end{remark}

\subsection{\ 3-forms}

\subsubsection{Topological Torsion and Topological Spin 3-forms}

The development above indicates that certain vectors of 3 forms are pertinent
to the theory of conserved currents. \ However, there are other important
3-forms that can be deduced from the single postulate that defines a Physical
Vacuum. \ These 3-forms have utilization in the understanding of\ both non
equilibrium thermodynamic phenomena and, remarkably enough, the concept of
Topological Spin.

Construct the (pair - even) Topological Torsion \cite{rmkada} \cite{vol4}
scalar of 3-forms, defined as
\begin{align}
\text{\textbf{Topological Torsion}: \ \ }  &  H=\left\langle A^{k}\right\vert
\symbol{94}\left\vert F^{k}\right\rangle \\
&  =A^{1}\symbol{94}F^{1}+A^{2}\symbol{94}F^{2}+A^{3}\symbol{94}F^{3}%
+A^{4}\symbol{94}F^{4}.
\end{align}
Note that the topological dimensions of $H$ are that of the Flux Quantum,
$\hbar/e$. \ The object $H$ is composed of 3-forms that are tensor invariants
relative to diffeomorphisms. \ 

Each of the four 3-form elements for a fixed index k, $A^{k}\symbol{94}F^{k}$,
if not zero, will indicate that the Pfaff Topological dimension of the
specific 1-form, $A^{k}$, is 3 or more. \ The idea is that it takes at least 3
independent functions (of the 4 space time variables) to describe the
topological features of the 1-form of given index, k. \ As equilibrium
thermodynamic phenomena require two independent functions at most
\cite{Caratheodory}, it follows that systems where the Pfaff Topological
dimension is 3 or more are non equilibrium systems.

Construct the (impair-odd)\ Topological Spin \cite{rmkAG} \cite{vol4} (pseudo)
scalar of 3-forms defined as
\begin{align}
\text{\textbf{Topological Spin}: \ \ \ }  &  S=\left\langle A^{m}\right\vert
\symbol{94}\left\vert G^{m}\right\rangle ,\\
&  =A^{1}\symbol{94}G^{1}+A^{2}\symbol{94}G^{2}+A^{3}\symbol{94}G^{3}%
+A^{4}\symbol{94}G^{4}.
\end{align}
For electromagnetic systems, the physical units of S are $\hslash$ (Planck's
constant of angular momentum). \ The concept of Topological Spin depends upon
the fact that the vector of excitation 2-forms, $\left\vert G\right\rangle ,$
is not zero. \ Consequently the coefficients of Affine torsion must not be
zero. \ "Affine Torsion" is necessary, but not sufficient to produce
Topological Spin. \ Examples of such properties will be given in Part 2.
\ Note that the individual 3-form components, $A\symbol{94}G$, of the
Topological Spin composite do not transform as a tensor invariant relative to
diffeomorphisms. \ Each term does depend on the Inverse Jacobian matrix of the
diffeomorphism. \ Each term has a pullback proportional to the adjoint of the
Jacobian matrix, divided by the determinant of the transformation. \ If the
diffeomorphisms are limited to elements of the special orthogonal (or
unitary)\ group\footnote{Such as SO3, or SO2.} then a dualism can be defined
between the vector of intensities and the vector of excitations. \ However,
this self (or anti-self) dualism is a very special case (assumed in the
Yang-Mills formalism, but not herein).

\subsubsection{Poincare Invariants}

Exterior differentiation of these two composite 3-forms, $H$ and $S$, produces
the Poincare invariants%
\begin{align}
dH  &  =\left\langle F^{k}\right\vert \symbol{94}\left\vert F^{k}\right\rangle
=\sum_{k}2(\mathbf{E\cdot B)}^{k}\ \ =\text{\textbf{Poincare II}}\\
dS  &  =\left\langle F^{k}\right\vert \symbol{94}\left\vert G^{k}\right\rangle
-\left\langle A^{k}\right\vert \symbol{94}\left\vert J^{k}\right\rangle
\ \ \ \ \ \ \ \ =\text{\textbf{Poincare I.}}\\
&  =\sum_{l}[\{(\mathbf{B\cdot H)}^{k}-(\mathbf{D\cdot E)}^{k}%
\}-\{(\mathbf{A\cdot J)}^{k}-(\rho\phi)^{k}\}].
\end{align}
The notation in terms of $\mathbf{E,B,D,H}$ is symbolic, and refers only to
the formal equivalence of the formulas to electromagnetic theory. \ If the
Poincare invariants vanish, the closed integrals of the closed 3-forms could
become topologically quantized through the concept of deRham period integrals.
\ There are many ways that this result could happen, due to the fact that
there are four (Yang Mills type) components to the excitation 2-forms,
$\left\vert G^{b}\right\rangle .$ Note that if there exists only 1-component
of $\left\vert G^{b}\right\rangle ,$ say $G^{4},\ $and it is closed,
$dG^{4}=0,$ then, formally, the integral over a closed 2D integration chain of
$G^{4}$ defines the Quantized Charge, e, of classical electromagnetic theory.%

\begin{equation}%
{\displaystyle\iint\limits_{closed}}
G^{4}\Rightarrow e.
\end{equation}

The multi-component Field intensity and Field excitation equations are
\textit{extensions} of a Yang Mills theory, as $\pm$ self duality is not
imposed on the system.

\subsection{Period Integrals and Topological Quantization.}

Although the main interest of this article is associated with the field
properties of the Physical Vacuum, a few words are appropriate about the
topological defect structures (that will be treated in more detail in another
article). \ Of specific interest are those topological structures represented
by closed, but not exact, differential forms. \ Such exterior differential
forms are homogeneous of degree zero expressions. \ Such closed structures can
lead to deRham period integrals \cite{rmkperiods}, whose values have rational
ratios, when the integration chain, z1, is also closed, and not equal to a
boundary. \ For example, the Flux Quantum of EM theory is given by the closed
integral of the (electrodynamic) 1-form of Action:%

\begin{align}
\text{Flux quantum}  &  =n%
{\displaystyle\oint_{z1}}
A\\
\text{ \ In domains where }F  &  =dA\Rightarrow0
\end{align}
This period integral is NOT\ dependent upon the electromagnetic field
intensities, $F$. \ Stokes theorem does not apply if the integration chain is
not a boundary. \ All integrals of exact forms over boundaries would yield zero.

As another example, the Period integral of quantized charge is given by the expression,%

\begin{align}
\text{Charge quantum}  &  =n%
{\textstyle\iint_{z2}}
G\text{ }\\
\text{ \ In domains where }J  &  =dG\Rightarrow0.
\end{align}
The integration is over a closed chain, z2, which is not a boundary. \ Again,
the quantized Period integral does not depend upon the charge current
3-form,\ $J$, and the expression is valid only in domains where $J\Rightarrow
0.$ \ These same properties of topological quantization are universal ideas
independent from the notation, or some topological refinement to a specific
types of physical systems. \ Note that the concept of the Charge quantum
depends upon the existence of $G$ which in turn depends upon the existence of
Affine Torsion of the Cartan Connection $[\mathbb{C}]$ based on $[\mathbf{B}]$.

\subsection{Quadratic Congruents and Metrics}

Starting from the existence of a Linear Basis Frame, it is remarkable that
symmetric properties of $\left[  \mathbb{B}\right]  $\ can be deduced in terms
of a quadratic congruence (see p. 36 in Turnbull and Aitken \cite{TurnAiken}).
The quadratic congruence is related to the concept of strain in elasticity
theory, and is quite different from the linear definition of matrix symmetries
in terms of the sum of a matrix $\left[  \mathbb{B}\right]  $ and its
transpose $\left[  \mathbb{B}\right]  ^{T}$. \ The algebraic quadratic
congruence will be used to define compatible symmetric (metric) qualities in
terms of the structure of the Basis Frames, $\left[  \mathbb{B}\right]  $:%
\begin{equation}
\left[  g\right]  =\left[  \mathbb{B}\right]  ^{T}\circ\left[  \eta\right]
\circ\left[  \mathbb{B}\right]  .
\end{equation}
The matrix $\left[  \eta\right]  $ is a (diagonal)\ Sylvestor signature matrix
whose elements are $\pm1.$ \ Recall that in projective geometry the congruence
transformation based upon $\left[  \mathbb{B}\right]  ^{T}$ defines a
correlation, where the similarity transformation based upon $\left[
\mathbb{B}\right]  ^{-1}$ defines a collineation. \ Note that the ubiquitous
choice of an orthonormal basis frame, where $\left[  \mathbb{B}\right]
^{T}=\left[  \mathbb{B}\right]  ^{-1},$ limits the topological generality of
the concept of a Physical Vacuum.

In a later section it will be demonstrated how the similarity invariants of
the Basis Frame find use in representing thermodynamic phase functions
appropriate to the Physical Vacuum.

\begin{remark}
Note that this \textit{quadratic} (multiplicative) symmetry property is not
the equivalent to the\ (additive) symmetry property defined by the
\textit{linear} sum of the matrix $\left[  \mathbb{B}\right]  $ and its
transpose. \ 
\end{remark}

However, from the Basis Frame, it is also possible to construct a topological
exterior differential system that defines a quadratic form of the law of
differential closure. \ It then follows that, for $d\left[  \eta\right]  =0,$
\begin{equation}
d\left[  g\right]  =[\widetilde{\mathbb{C}}_{r}]\circ\left[  g\right]
+\left[  g\right]  \circ\left[  \mathbb{C}_{r}\right]  .\text{ }%
\end{equation}
This deductive result leads to the metricity condition for a Physical Vacuum,
in terms of the right Cartan connection $\left[  \mathbb{C}_{r}\right]  $:%
\begin{equation}
\text{\textbf{Metricity condition:} \ }d\left[  g\right]  -[\widetilde
{\mathbb{C}}_{r}]\circ\left[  g\right]  -\left[  g\right]  \circ\left[
\mathbb{C}_{r}\right]  \Rightarrow0.
\end{equation}
This equation is an exterior differential system, and therefor defines
topological properties.

Compute the Christoffel connection, and its matrix of 1-forms, $\left[
\Gamma\right]  ,$\ from the quadratic "metric" matrix $\left[  g\right]  ,$
using the Levi-Civita-Christoffel formulas.%
\begin{align}
\text{Coefficients}  &  :\text{Christoffel Connection}\\
\Gamma_{ac}^{b}(\xi^{c})  &  =g^{be}\{\partial g_{ce}/\partial\xi^{a}+\partial
g_{ea}/\partial\xi^{c}-\partial g_{ac}/\partial\xi^{e}\},\\
\text{ }\left[  \Gamma\right]   &  =\left[  \Gamma_{ac}^{b}dy^{c}\right]
\text{ \ as a matrix of 1-forms}%
\end{align}
The Christoffel Connection also satisfies the metricity condition,%

\begin{equation}
\text{\textbf{Metricity condition:} \ }d\left[  g\right]  -\left[
\Gamma\right]  \circ\left[  g\right]  -\left[  g\right]  \circ\left[
\Gamma\right]  \Rightarrow0.
\end{equation}

\subsubsection{Decomposition of the Cartan connection}

The linear properties of matrices permits the Cartan Connection matrix
$\left[  \mathbb{C}\right]  $ to be decomposed into a term, $\left[
\Gamma\right]  ,$ based on its quadratic congruent symmetries, and a residue
$\left[  \mathbb{T}\right]  $ that will contain any asymmetries due to non
zero Affine Torsion coefficients. \ Decompose the Cartan Connection matrix of
1-forms as follows:
\begin{equation}
\left[  \mathbb{C}\right]  =\left[  \mathbf{\Gamma}\right]  +\left[
\mathbb{T}\right]  \label{decomp}%
\end{equation}
As the topological metricity condition is automatically satisfied for
Christoffel Connection $\left[  \mathbf{\Gamma}\right]  $, as well as for the
Cartan Connection, it must also be satisfied for\ residue matrix of 1-forms,
$\left[  \mathbb{T}\right]  $. \ 

\subsubsection{Matrices of Curvature 2-forms}

Construct the matrix of Cartan Curvature 2-forms, $\left[  \mathbf{\Phi
}\right]  ,\,$\ derived from the second exterior differentiation of the Basis
Frame, $dd\left[  \mathbb{B}\right]  ,$ and based on the Cartan Connection,
$\left[  \mathbb{C}\right]  :$
\begin{align}
dd\left[  \mathbb{B}\right]   &  =d\left[  \mathbb{B}\right]  \circ\left[
\mathbb{C}\right]  +\left[  \mathbb{B}\right]  \circ d\left[  \mathbb{C}%
\right]  ,\\
&  =\left[  \mathbb{B}\right]  \circ\{d\left[  \mathbb{C}\right]  +\left[
\mathbb{C}\right]  \symbol{94}\left[  \mathbb{C}\right]  \},\\
&  =\left[  \mathbb{B}\right]  \circ\left[  \mathbf{\Phi}\right]
\Rightarrow0,\\
\text{Curvature 2-forms }\left[  \mathbf{\Phi}\right]   &  =\{d\left[
\mathbb{C}\right]  +\left[  \mathbb{C}\right]  \symbol{94}\left[
\mathbb{C}\right]  \},\label{curv2f}\\
\left[  \mathbf{\Phi}\right]   &  =dd\left[  \mathbb{B}\right]  \Rightarrow0.
\end{align}
The zero result, based upon the Poincare lemma,\ $dd\left[  \mathbb{B}\right]
\Rightarrow0,$ requires C2 differentiability of the Basis Frame functions.

The formula given by eq(\ref{curv2f}) is called Cartan's second (sometimes the
first) structural equation and is based upon a matrix of curvature 2-forms.
\ The formula can be applied to arbitrary connections (not defined in terms of
the Basis Frame of the Physical Vacuum) and will yield non zero values.
\ Similar remarks may be made about Cartan's first equation of structure based
on Cartan's vector of Torsion 2-forms given by eq(\ref{tors2f}). \ Cartan's
matrix of curvature 2-forms, and Cartan's vector of Torsion 2-forms are both
zero for the "Physical Vacuum". \ The "Affine" torsion coefficients of the
Cartan Matrix need not be zero for the "Physical Vacuum".

\subsubsection{Bianchi Identities and Matrices of 3-forms.}

Note that exterior differentiation of the Cartan structure matrix of curvature
2-forms is equivalent to the Bianchi identity:%
\begin{align}
\left[  d\mathbf{\Phi}\right]  +\left[  d\mathbb{C}\right]  \symbol{94}\left[
\mathbb{C}\right]  -\left[  \mathbb{C}\right]  \symbol{94}\left[
d\mathbb{C}\right]   &  =\\
\left[  d\mathbf{\Phi}\right]  +\left[  \mathbf{\Phi}\right]  \symbol{94}%
\left[  \mathbb{C}\right]  -\left[  \mathbb{C}\right]  \symbol{94}\left[
\mathbf{\Phi}\right]   &  \Rightarrow0.
\end{align}
This concept of a Bianchi identity is valid for all forms of the Cartan
structure equations. \ The Bianchi statements are essentially definitions of
cohomology, in that the difference between two non-exact p-forms is equal to a
perfect differential (an exterior differential system). \ In this case the
Bianchi identity describes the cohomology established by two matrices of
3-forms, $\left[  J2\right]  -\left[  J1\right]  $.%
\begin{align}
\left[  J2\right]  -\left[  J1\right]   &  =\left[  d\mathbf{\Phi}\right]  ,\\
\text{where }\left[  J1\right]   &  =\left[  d\mathbb{C}\right]
\symbol{94}\left[  \mathbb{C}\right] \\
\text{and }\left[  J2\right]   &  =\left[  \mathbb{C}\right]  \symbol{94}%
\left[  d\mathbb{C}\right]  .
\end{align}

\subsection{The Higgs vector of zero forms (Internal Energy)}

Once again consider the Lie differential with respect to a direction field $V
$, operating on the formula for differential closure%
\begin{align}
L_{(V)}(\left[  \mathbb{B}(y)\right]  \circ\left\vert dy^{a}\right\rangle )\
&  =L_{(V)}(\left\vert A^{a}\right\rangle )=i(V)d\left\vert A^{a}\right\rangle
+d(i(V)\left\vert A^{a}\right\rangle )\\
&  =i(V)\left\vert F^{a}\right\rangle +d(i(V)\left\vert A^{a}\right\rangle
)=\\
&  =\left\vert W^{a}\right\rangle +d\left\vert h^{a}\right\rangle .
\end{align}
From Koszul's theorem, $\left\vert W^{a}\right\rangle =i(V)d\left\vert
A^{a}\right\rangle $ is a covariant differential based on some (abstract)
connection (for each $a).$ \ Hence, the difference between the Lie
differential and the Covariant differential is the exact term,
$d(i(V)\left\vert A^{a}\right\rangle ):$%
\begin{equation}
L_{(V)}(\left\vert A^{a}\right\rangle )-i(V)d\left\vert A^{a}\right\rangle
=d(i(V)\left\vert A^{a}\right\rangle )=d\left\vert h^{a}\right\rangle .
\end{equation}
This equation is another statement of Cohomology, another exterior
differential system, where the difference of two non-exact objects is an exact differential.

From the topological formulation of thermodynamics \cite{vol1} in terms of
Cartan's magic formula \cite{Marsden},
\begin{align}
\text{ Cartan's Magic Formula }L_{(\rho\mathbf{V}_{4})}A  &  =i(\rho
\mathbf{V}_{4})dA+d(i(\rho\mathbf{V}_{4})A)\\
\text{First Law }  &  :W+dU=Q,\\
\text{Inexact Heat 1-form\ \ }Q  &  =W+dU=L_{(\rho\mathbf{V}_{4})}A\\
\text{Inexact Work 1-form\ }W  &  =i(\rho\mathbf{V}_{4})dA,\\
\text{Internal Energy \ }U  &  =i(\rho\mathbf{V}_{4})A,
\end{align}
Now consider particular process paths (defined by the directional field
$\rho\mathbf{V}_{4})$, and deduce that in the direction of the process path
\begin{align}
\text{ }i(\rho\mathbf{V}_{4})W  &  =0,\text{ }\\
\text{Work }  &  :\text{ is transversal;}\\
i(\rho\mathbf{V}_{4})Q  &  =i(\rho\mathbf{V}_{4})dU\text{ }\neq0\\
\text{ Heat }  &  :\text{is not transversal;}\\
\text{but if \ }i(\rho\mathbf{V}_{4})Q  &  =0,\text{ }\\
\text{the process }  &  :\text{is adiabatic.}%
\end{align}

It is the non-adiabatic components of a thermodynamic process that indicate
that there is a change of internal energy and hence an inertial force in the
direction of a process. \ This implies that the non-adiabatic processes are
inertial effects, and could be related to changes in mass.

Now to paraphrase a statements and ideas from Mason and Woodhouse, (see p. 49
\ \cite{MW}) and \cite{atiyah} :

"... then there is a Higgs field.... which measures the difference between the
Covariant differential along $V$ and the Lie differential along $V$."

It becomes apparent that the
\begin{align}
\left\vert W^{a}\right\rangle  &  =\text{Vector of Work 1-forms.
(transversal)}\\
\left\vert h^{a}\right\rangle  &  =\text{Higgs potential as vector of 0-forms
(Internal Energy)}\\
d\left\vert h^{a}\right\rangle  &  =\text{Higgs vector of 1-forms.}\\
i(V)d\left\vert h^{a}\right\rangle  &  =\text{vector of longitudinal inertial
accelerations}\\
&  =\text{non adiabatic components of a process}%
\end{align}
The method of the "Physical Vacuum" and its sole assumption\ leads to inertial
properties and the Higgs field, all from a topological perspective and without
"quantum" fluctuations.\ 

\subsection{A Strong Equivalence Principle}

At this point, there has been no indication that the problem being
investigated has anything to do with the Gravitational Field. \ The gravity
issue is to be encoded into how the quadratic congruent symmetries of $\left[
\mathbb{B}\right]  ,$ and its topological group structures, are established.
\ In general, different choices for the group structure of the Basis Frame
will strongly influence the application to any particular physical system of
fields and particles. \ 

Without the Einstein Ansatz, it appears that the concept of a Physical Vacuum
can lead to a Strong Equivalence principle. \ Substitute $\left[
\Gamma\right]  +\left[  \mathbb{T}\right]  $ for $\left[  \mathbb{C}\right]  $
in the definition of the matrix of curvature 2-forms, and recall that for the
Physical Vacuum the Cartan matrix of curvature 2-forms, $\left[  \mathbf{\Phi
}\right]  ,$ is zero.%
\begin{align}
\left[  \mathbf{\Phi}_{\mathbb{C}}\right]   &  =\{d\left[  \mathbb{C}\right]
+\left[  \mathbb{C}\right]  \symbol{94}\left[  \mathbb{C}\right]
\}\Rightarrow0,\\
&  =\{d(\left[  \Gamma\right]  +\left[  \mathbb{T}\right]  )+(\left[
\Gamma\right]  +\left[  \mathbb{T}\right]  )\symbol{94}(\left[  \Gamma\right]
+\left[  \mathbb{T}\right]  )\}\\
&  =\{d[\Gamma]+[\Gamma]\symbol{94}[\Gamma]\}+\{\left[  \mathbb{T}\right]
\symbol{94}\left[  \Gamma\right]  +\left[  \Gamma\right]  \symbol{94}\left[
\mathbb{T}\right]  \}+\{d[\mathbb{T}]+[\mathbb{T}]\symbol{94}\mathbb{[T}]\},
\end{align}

Separate the matrices of 2-forms into the metric based (Christoffel) curvature
2-forms, defined as
\begin{equation}
\left[  \mathbf{\Phi}_{\mathbf{\Gamma}}\right]  =\{d\left[  \Gamma\right]
+\left[  \Gamma\right]  \symbol{94}\left[  \Gamma\right]  \}=\left[
Field\ metric\ 2-forms\right]  ,
\end{equation}
and the remainder, defined as
\begin{align}
\left[  -\mathbf{\Phi}_{Inertial}\right]   &  =\left[  \mathbf{\Phi
}_{\mathbb{C}}\right]  -\left[  \mathbf{\Phi}_{\mathbf{\Gamma}}\right] \\
&  =\{\left[  \mathbb{T}\right]  \symbol{94}\left[  \Gamma\right]  +\left[
\Gamma\right]  \symbol{94}\left[  \mathbb{T}\right]  \}+\{d[\mathbb{T}%
]+[\mathbb{T}]\symbol{94}\mathbb{[T}]\}\\
&  =\{interaction\_2-forms\}\ \ +\ \{\left[  \mathbf{\Phi}_{\mathbb{T}%
}\right]  \}
\end{align}
$.$ The decomposition leads to the strong equivalence equation,%
\begin{align}
\text{Principle of }  &  :\text{Strong Equivalence }\\
\left[  \text{Metric Field curvature 2-forms}\right]   &  =\left[
\text{Inertial curvature 2-forms}\right]  ,\\
\left[  \mathbf{\Phi}_{\mathbf{\Gamma}}\right]   &  =\left[  -\mathbf{\Phi
}_{Inertial}\right]
\end{align}

\subsection{The Source of electromagnetic Charge and Spin}

A number of years ago, it became apparent to me that the origin of charge was
to be associated with topological structures of space time\ \cite{rmkcharge}.
\ The concepts exploited in this study assumed the existence of an impair
2-form of field intensities $G.$ \ Based on the development above, it can be stated:

\begin{remark}
The theory of a Physical Vacuum asserts that the existence of charge is
dependent upon the Affine Torsion of the Cartan connection $[\mathbb{C}]$.
\end{remark}

This result is based upon the\ formal correspondence between equations of
electromagnetic field excitations $\left\vert G\right\rangle $ and the "Affine
torsion" coefficients (not the Cartan torsion coefficients) deduced for the
Cartan Connection matrix of 1-forms $\left[  \mathbb{C}\right]  $.%

\begin{equation}
\text{Excitation 2-forms }\left\vert G\right\rangle =\left[  \mathbb{C}%
\right]  \symbol{94}\left\vert dy\right\rangle \text{ Affine Torsion 2-forms.}%
\end{equation}
\ The impair 2 forms that compose the elements of $\left\vert G\right\rangle $
formally define the field excitations in terms of the coefficients of "Affine
torsion". \ The closed period integrals of those (closed but not exact)
components of $\left\vert G\right\rangle ,$ which are homogeneous of degree 0,
lead to the deRham integrals with rational, quantized, ratios. \ Such
excitation 2-forms $G$ do not exist if the coefficients of Affine torsion are zero.

The topological features of the Physical Vacuum are determined by the
structural properties of the Basis Frames, $\left[  \mathbb{B}\right]  ,$ and
the derived Cartan Connection matrix of 1-forms, $\left[  \mathbb{C}\right]
.$%

\begin{align*}
&  \text{\textbf{\ \ \ \ \ \ \ \ \ \ \ \ \ \ The Topological Structure of the
Physical Vacuum}}\\
&
\text{\textbf{\ \ \ \ \ \ \ \ \ \ \ \ \ \ \ \ \ \ \ \ \ \ \ \ \ \ \ \ \ \ \ \ \ \ \ \ \ in
terms of }}\left[  \mathbb{B}\right]  \text{ \textbf{and} }\left[
\mathbb{C}\right] \\
&  \left[
\begin{array}
[c]{ccc}%
\text{\textbf{Mass}} &
\begin{array}
[c]{c}%
\text{Non Zero Cartan Curvature of}\\
\text{of }\left[  \Gamma\right]  \text{ =}\left[  \mathbb{B}\right]  ^{T}%
\circ\left[  \eta\right]  \circ\left[  \mathbb{B}\right]
\end{array}
&
\begin{array}
[c]{c}%
\text{Cartan Curvature of }\left[  \mathbb{C}\right] \\
\text{is ZERO}%
\end{array}
\\
\text{\textbf{Charge} } &
\begin{array}
[c]{c}%
\text{Non Zero Affine Torsion of }\left[  \mathbb{C}\right] \\
\left\vert G\right\rangle =\left[  \mathbb{C}\right]  \symbol{94}\left\vert
dy\right\rangle
\end{array}
&
\begin{array}
[c]{c}%
\text{Cartan Torsion of }\left[  \mathbb{C}\right] \\
\text{is ZERO}%
\end{array}
\end{array}
\right]
\end{align*}

Where the presence of mass is recognized in terms of the Riemannian curvature
of the quadratic congruences of a Physical Vacuum, the presence of charge is
recognized in terms of the coefficients of Affine Torsion of a Physical Vacuum.

\section{Remarks}

This universal set ideas enumerated in Part I startles me. \ There is only ONE
fundamental assumption, and the rest of the concepts are derived, following
the rules of the Cartan exterior calculus. \ The results appear to be
universal rules. \ Under topological refinements, the rules are specialized
into domains that are recognizable as having the features of the four forces
of physics. \ Not only do the long range concepts (fields) of gravity and
electromagnetism have the same base, but so also do the short range concepts
(fields) of the nuclear and weak force have the same base. \ Earlier, related,
thoughts about the topological and differential geometric ideas associated
with the four forces appeared in \cite{rmksubmersive}. \ Now the topological
theory of a Physical Vacuum as presented above re-enforces the earlier work.
\ In addition, particulate concepts also appear from the same fundamental
postulate of a Physical Vacuum. \ They appear as topologically coherent defect
structures in the fields. \ Quantization occurs in a topological manner from
the deRham theorems as period integrals. \ The quantized topologically
coherent structures form the basis of macroscopic quantum states.

\part{Examples}

\section{Example 1. \ \textbf{The Schwarzschild Metric embedded in a Basis
Frame,\ [}$\mathbb{B}$\textbf{], as a 10 parameter subgroup of an affine
group.}}

\subsection{The Metric - a Quadratic Congruent symmetry}

The algebra of a quadratic congruence can be used to deduce the metric
properties of a given Basis Frame. \ These deduced metric features may be used
to construct a "Christoffel" or metric compatible connection, different from
the Cartan Connection. \ The Christoffel connection constructed from the
quadratic congruence of the Basis Frame may or may not generate a "Riemannian"
curvature. \ By working backwards, this example will demonstrate how to
construct the Basis Frame, given a metric. \ The method is algebraic and
exceptionally simple for all metrics that represent a 3+1 division of
space-time. \ The important result is that given a Basis Frame of a given
matrix group structure, a metric and a compatible Christoffel Connection can
be deduced.

\begin{remark}
All 3+1 metric structures are to be associated with the 10 parameter subgroup
of the 13 parameter Affine groups. \ 
\end{remark}

In this example, it will be demonstrated how the isotropic form of the
Schwarzschild metric can be incorporated into the Basis Frame for a Physical
Vacuum, $\left[  \mathbb{B}\right]  $. \ The technique is easily extendable
for diagonal metrics. \ However, the symmetry properties of the Cartan
Connection are not limited to metrics of the "gravitational" type. \ Once the
Schwarzschild metric is embedded in to the Basis Frame, then the universal
methods described above will be applied to the representative Basis Frame, and
each important result will be evaluated.

The isotropic Schwarzschild metric is a diagonal metric of the form,%

\begin{align}
(\delta s)^{2}  &  =-(1+m/2r)^{4}\{(dx)^{2}+(dy)^{2}+(dz)^{2}\}+\frac
{(2-m/r)^{2}}{(2+m/r)^{2}}(dt)^{2}\\
&  =-(\alpha)^{2}\{(dx)^{2}+(dy)^{2}+(dz)^{2}\}+(\beta)^{2}(dt)^{2}\\
\text{with }r  &  =\sqrt{(x)^{2}+(y)^{2}+(z)^{2}},
\end{align}
As Eddington \cite{EddingtonREL} points out, the isotropic form is palatable
with the idea that the speed of light is equivalent in any direction. \ That
is not true for the non-isotropic Schwarzschild metric, where transverse and
longitudinal null geodesics do not have the same speed. \ \ 

For the isotropic Schwarzschild example, the metric $\left[  g_{jk}\right]
$\ can be constructed from the triple matrix product:
\begin{align}
\left[  g_{jk}\right]   &  =[\widetilde{f}]\circ\left[  \eta\right]
\circ\left[  f\right]  ,\\
\text{where }f  &  =\left[
\begin{array}
[c]{cccc}%
\alpha & 0 & 0 & 0\\
0 & \alpha & 0 & 0\\
0 & 0 & \alpha & 0\\
0 & 0 & 0 & \beta
\end{array}
\right]  ,\\
\text{and \ \ \ }\alpha &  =(1+m/2r)^{2}\ =(\gamma/2r)^{2},\ \ \ \ \beta
=\frac{(2-m/r)}{(2+m/r)}\ =\delta/\gamma,\\
\text{and \ \ \ }\eta &  =\left[
\begin{array}
[c]{cccc}%
-1 & 0 & 0 & 0\\
0 & -1 & 0 & 0\\
0 & 0 & -1 & 0\\
0 & 0 & 0 & 1
\end{array}
\right]  .
\end{align}
At first glance it would appear that the Schwarzschild metric forms a
quadratic form constructed from the congruence of a 4 parameter (diagonal)
matrix. \ It is not obvious that this may be a special case of a 10 parameter
group with a fixed point. \ In order to admit the 10 parameter Poincare group
(which is related to the Lorentz group), a map from spherical 3+1 space to
Cartesian 3+1 space will be perturbed by the matrix $[f].$

\subsubsection{The Diffeomorphic Jacobian Basis Frame}

At first, consider the diffeomorphic map $\phi^{k}$ from spherical\ 3+1 space
to Cartesian 3+1 coordinates:
\begin{align}
\{y^{a}\}  &  =\{r,\theta,\varphi,\tau\}\Rightarrow\{x^{k}\}=\{x,y,z,t]\\
\phi^{k}  &  :[r\sin(\theta)\cos(\varphi),r\sin(\theta)\sin(\varphi
),rcos(\theta),\tau]\Rightarrow\lbrack x,y,z,t]\\
\{dy^{a}\}  &  =\{dr,d\theta,d\phi.d\tau\}.
\end{align}
The Jacobian of the diffeomorphic map $\phi^{k}$ can be utilized as an
integrable Basis Frame matrix\ $\left[  \mathbb{B}\right]  $ which is an
element of the 10 parameter F-Affine group (The Affine subgroup with a fixed point):%

\begin{equation}
\left[  \mathbb{B}\right]  =\left[
\begin{array}
[c]{cccc}%
\sin(\theta)\cos(\varphi) & r\cos(\theta)\cos(\varphi) & -r\sin(\theta
)\sin(\varphi) & 0\\
\sin(\theta)\sin(\varphi) & r\cos(\theta)\sin(\varphi) & r\sin(\theta
)\cos(\varphi) & 0\\
\cos(\theta) & -r\sin(\theta) & 0 & 0\\
0 & 0 & 0 & 1
\end{array}
\right]  .
\end{equation}
\bigskip

The infinitesimal mapping formula based on $\left[  \mathbb{B}\right]  $ yields%

\begin{equation}
\left[  \mathbb{B}_{a}^{k}(y)\right]  \circ\left\vert
\begin{array}
[c]{c}%
dr\\
d\theta\\
d\varphi\\
dt
\end{array}
\right\rangle \ \Rightarrow\ \left\vert \sigma^{k}\right\rangle =\left\vert
\begin{array}
[c]{c}%
dx\\
dy\\
dz\\
dt
\end{array}
\right\rangle ,
\end{equation}
such that all of the 1-forms $\left\vert \sigma^{k}\right\rangle $ are exact differentials.

\subsubsection{The Perturbed Basis Frame with a Congruent Symmetry}

\begin{theorem}
\ \ \textit{The effects of a diagonal metric }$\left[  g_{jk}\right]
$\textit{\ can be absorbed into a re-definition of the Frame matrix:}%
\begin{equation}
\lbrack\widehat{\mathbb{B}}]=\left[  f\right]  \circ\lbrack\mathbb{B}].\
\end{equation}

\end{theorem}

The integrable Jacobian Basis Frame matrix given above will be perturbed by
multiplication on the left by the diagonal matrix, $\left[  f\right]  .$ \ The
perturbed Basis Frame becomes%

\begin{align}
\lbrack\widehat{\mathbb{B}}]  &  =\left[  f\right]  \circ\lbrack
\mathbb{B}]\text{ \ the Schwarzschild Cartan Basis Frame.}\\
&  =\left[
\begin{array}
[c]{cccc}%
\sin(\theta)\cos(\varphi)\gamma^{2}/4r^{2} & \cos(\theta)\cos(\varphi
)\gamma^{2}/4r & -\sin(\theta)\sin(\varphi)\gamma^{2}/4r & 0\\
\sin(\theta)\sin(\varphi)\gamma^{2}/4r^{2} & \cos(\theta)\sin(\varphi
)\gamma^{2}/4r & \sin(\theta)\cos(\varphi)\gamma^{2}/4r & 0\\
\cos(\theta)\gamma^{2}/4r^{2} & -\sin(\theta)\gamma^{2}/4r & 0 & 0\\
0 & 0 & 0 & \delta/\gamma
\end{array}
\right]  .\ \\
\gamma &  =(2r+m),\ \ \ \ \delta=(2r-m)
\end{align}
Use of the congruent pullback formula based on the perturbed Basis Frame,
$[\widehat{\mathbb{B}}]$, yields,%

\begin{align}
\left[  g_{jk}\right]   &  =[\widehat{\mathbb{B}}_{transpose}]\circ\eta
\circ\lbrack\widehat{\mathbb{B}}],\\
\left[  g_{jk}\right]   &  =\left[
\begin{array}
[c]{cccc}%
-(\gamma^{2}/4r^{2})^{2} & 0 & 0 & 0\\
0 & -(\gamma^{2}/4r)^{2} & 0 & 0\\
0 & 0 & -(\gamma^{2}/4r)^{2}\sin^{2}(\theta) & 0\\
0 & 0 & 0 & +(\delta/\gamma)^{2}%
\end{array}
\right]  ,\\
\gamma &  =(2r+m),\ \ \ \ \delta=(2r-m)
\end{align}
which agrees with formula given above for the isotropic Schwarzschild metric
in spherical coordinates. \ It actually includes a more general idea, for the
coefficients, $\alpha,$and $\beta,$ can be dependent upon both $r$ and $\tau$. \ 

The infinitesimal mapping formula based on $[\widehat{\mathbb{B}}]$ yields%

\begin{equation}
\left[  \mathbb{B}_{a}^{k}(y)\right]  \circ\left\vert
\begin{array}
[c]{c}%
dr\\
d\theta\\
d\varphi\\
dt
\end{array}
\right\rangle \ \Rightarrow\ \left\vert \sigma^{k}\right\rangle =\left\vert
\begin{array}
[c]{c}%
(\gamma^{2}/4r^{2})dx\\
(\gamma^{2}/4r^{2})dy\\
(\gamma^{2}/4r^{2})dz\\
(\delta/\gamma)dt
\end{array}
\right\rangle ,
\end{equation}
such that all of the 1-forms $\left\vert \sigma^{k}\right\rangle $ are NOT
exact differentials.

\subsection{The Schwarzschild-Cartan connection.}

The Schwarzschild-Cartan (right) Connection $[\widehat{\mathbb{C}}],$ as a
matrix of 1-forms relative to the perturbed Basis Frame $[\widehat{\mathbb{B}%
}],$ becomes%

\begin{align}
\lbrack\widehat{\mathbb{C}}]  &  =[\widehat{\mathbb{B}^{-1}}]\circ
d[\widehat{\mathbb{B}}],\\
\lbrack\widehat{\mathbb{C}}]  &  =\left[
\begin{array}
[c]{cccc}%
-2mdr/r\gamma & -rd\theta & \sin^{2}(\theta)rd\phi & 0\\
d\theta/r & \delta dr/\gamma & -\cos(\theta)\sin(\theta)d\phi & 0\\
d\phi/r & \cot(\theta)d\phi & \cot(\theta)d\theta+\delta dr/\gamma & 0\\
0 & 0 & 0 & 4mdr/(\gamma\delta)
\end{array}
\right]  .\\
\gamma &  =(2r+m),\ \ \ \ \delta=(2r-m)
\end{align}

\begin{center}
\textbf{Perturbed Cartan Connection\bigskip\ from Maple}
\end{center}

It is apparent that the Cartan Connection matrix of 1-forms is again a member
of the 10 parameter matrix group, a subgroup of the 13 parameter affine matrix group.

\subsection{Vectors of Torsion 2-forms}

Surprisingly, for the perturbed Basis Frame $[\widehat{\mathbb{B}}]$\ which
contains a the square root of a congruent metric field of a massive object,
the vector of excitation torsion 2-forms, based on the 10 parameter affine
subgroup with a fixed point, is not zero, and can be evaluated as:%

\begin{align}
\widehat{\left\vert G\right\rangle }  &  =[\widehat{\mathbb{C}}]\symbol{94}%
\left\vert dy^{a}\right\rangle \text{ \ }\\
\text{Torsion 2-forms }  &  :\text{\ of the Affine subgroup with a fixed
point}\\
\widehat{\left\vert G\right\rangle }  &  =\left\vert
\begin{array}
[c]{c}%
0\\
(2m/r\gamma)(d\theta\symbol{94}dr)\\
(2m/r\gamma)(d\phi\symbol{94}dr)\\
(4m/\gamma\delta)(dr\symbol{94}d\tau)
\end{array}
\right\rangle \text{ "Schwarzschild Excitations" }\\
\gamma &  =(2r+m),\ \ \ \ \delta=(2r-m)
\end{align}
The unexpected result is that the isotropic Schwarzschild metric admits
coefficients of "affine torsion" relative to the Cartan Connection matrix,
$[\widehat{\mathbb{C}}]$.

Similarly the vector of 2-form of field intensities $\widehat{\left\vert
F\right\rangle }$ can be evaluated in terms of the perturbed Basis Frame as:%
\begin{align}
\widehat{\left\vert F\right\rangle }  &  =d([\widehat{\mathbb{B}}%
]\circ\left\vert dy^{a}\right\rangle )=\text{ }\left\vert dA^{k}\right\rangle
\text{ \ \ \ \ \ "Schwarzschild Intensities" }\\
&  :\text{Intensity 2-forms of the Affine subgroup with a fixed point}\\
\widehat{\left\vert F\right\rangle }  &  =\left\vert
\begin{array}
[c]{c}%
+(m\gamma/2r^{2})\{(\sin(\phi)\cos(\theta)d\theta\symbol{94}dr)-(\sin
(\theta)\cos(\phi)d\phi\symbol{94}dr)\\
+(m\gamma/2r^{2})\{(\sin(\phi)\cos(\theta)d\theta\symbol{94}dr)+(\sin
(\theta)\cos(\phi)d\phi\symbol{94}dr)\\
-(m\gamma/2r^{2})(\sin(\theta)d\theta\symbol{94}dr)\\
(4m/\gamma^{2})(dr\symbol{94}d\tau)
\end{array}
\right\rangle .\\
\gamma &  =(2r+m),\ \ \ \ \delta=(2r-m)
\end{align}

The constitutive map relating the field intensities and the field excitations,%

\begin{equation}
\widehat{\left\vert G\right\rangle }=[\widehat{\mathbb{B}}]^{-1}\circ
\widehat{\left\vert F\right\rangle },
\end{equation}
is determined by the inverse of the perturbed Basis Frame, $[\widehat
{\mathbb{B}}]^{-1}$ :

\begin{center}%
\begin{align}
&  \text{\textbf{Schwarzschild Constitutive map from Maple}}\\
\lbrack\widehat{\mathbb{B}}]^{-1}  &  =4r/\gamma^{2}\left[
\begin{array}
[c]{cccc}%
r\sin(\theta)\cos(\phi) & r\sin(\theta)\sin(\phi) & \cos(\theta) & 0\\
\cos(\theta)\cos(\phi) & \cos(\theta)\sin(\phi) & -\sin(\theta) & 0\\
-\frac{\sin(\phi)}{\sin(\theta)} & \frac{\cos(\phi)}{\sin(\theta)} & 0 & 0\\
0 & 0 & 0 & \gamma^{3}/4r\delta
\end{array}
\right] \\
\gamma &  =(2r+m),\ \ \ \ \delta=(2r-m)
\end{align}

\end{center}

\subsubsection{Vectors of 3-forms}

\ The exterior derivative of the vector of excitations is zero, hence there
are no current 3-forms for the perturbed Basis Frame that encodes the
Schwarzschild metric as a Congruent symmetry:%

\begin{equation}
\text{Charge Current 3-form \ \ }\left\vert J\right\rangle =d\left\vert
G\right\rangle =0.
\end{equation}

The Topological Torsion for the Schwarzschild example vanishes:
$H=\left\langle A\right\vert \symbol{94}\left\vert F\right\rangle
$\ $\Rightarrow0.\ $The implication is that the system is of Pfaff dimension 2
(and therefor is an equilibrium thermodynamic system).

Both Poincare invariant 4-forms vanish, but the Topological Spin 3-form is NOT\ zero.%

\begin{align}
\text{Topological Spin 3-form \ \ \ \ }S  &  =\left\langle A\right\vert
\symbol{94}\left\vert G\right\rangle \ \ \ \ \ dA=0.\\
&  =(-m\gamma\sin(\theta)/2r^{2})\{\cos(\phi)+1)(dr\symbol{94}d\theta
\symbol{94}d\phi)\}.
\end{align}
The Spin 3-form depends upon the "mass" coefficient, $m$, and the 2-forms of
Affine torsion.

\subsubsection{The three Connection matrices}

The three matrices of Connection 1-forms are presented below for each
(perturbed) connection, $[\Gamma],~[\mathbb{C}],~[\mathbb{T]}$%

\begin{align}
\lbrack\widehat{\mathbf{\Gamma}}]  &  =\left[
\begin{array}
[c]{cccc}%
-2mdr/r\gamma & -\delta rd\theta/\gamma & \delta\sin^{2}(\theta)rd\phi/\gamma
& 64\delta md\tau/\gamma^{7}\\
\delta d\theta/r\gamma & \delta dr/r\gamma & -\cos(\theta)\sin(\theta)d\phi &
0\\
\delta d\phi/r\gamma & \cot(\theta)d\phi & \cot(\theta)d\theta+\delta
dr/\gamma & 0\\
4md\tau/(\gamma\delta) & 0 & 0 & 4mdr/(\gamma\delta)
\end{array}
\right]  .\\
\lbrack\widehat{\mathbb{C}}]  &  =\left[
\begin{array}
[c]{cccc}%
-2mdr/r\gamma & -rd\theta & \sin^{2}(\theta)rd\phi & 0\\
d\theta/r & \delta dr/\gamma & -\cos(\theta)\sin(\theta)d\phi & 0\\
d\phi/r & \cot(\theta)d\phi & \cot(\theta)d\theta+\delta dr/\gamma & 0\\
0 & 0 & 0 & 4mdr/(\gamma\delta)
\end{array}
\right]  .\\
\lbrack\widehat{\mathbb{T}}]  &  =\left[
\begin{array}
[c]{cccc}%
0 & -2mrd\theta/\gamma & 2m\sin^{2}(\theta)rd\phi/\gamma & -64mr^{4}\delta
d\tau/\gamma^{7}\\
2md\theta/r\gamma & 0 & 0 & 0\\
4md\phi/r\gamma & 0 & 0 & 0\\
-4md\tau/\delta\gamma & 0 & 0 & 0
\end{array}
\right]  .\\
\gamma &  =(2r+m),\ \ \ \ \delta=(2r-m)
\end{align}

\begin{center}
\textbf{Schwarzschild Perturbed Connections}
\end{center}

The matrix of (metric) curvature 2-forms, $\left[  \Phi_{\Gamma}\right]  ,$
based on the formula%

\begin{equation}
\left[  \Phi_{\Gamma}\right]  =d[\Gamma]+[\Gamma]\symbol{94}[\Gamma],
\end{equation}

\begin{center}
is computed to be:%
\begin{align}
&
\text{\textbf{Curvature\ 2-forms\ for\ the\ Schwarzschild\ Christoffel\ Connection}%
}\\
\lbrack\Phi_{\Gamma}]  &  =4m/\gamma^{2}\left[
\begin{array}
[c]{cccc}%
0 & -rdr\symbol{94}d\theta & r\sin^{2}(\theta)dr\symbol{94}d\phi &
-32r^{4}dr\symbol{94}d\tau/\gamma^{6}\\
2mdr\symbol{94}d\theta/r\gamma & 0 & -2\sin^{2}(\theta)d\theta\symbol{94}d\phi
& 16\delta^{2}r^{3}d\theta\symbol{94}d\tau/\gamma^{6}\\
4mdr\symbol{94}d\phi/r\gamma & -2rd\theta\symbol{94}d\phi & 0 & 16\delta
^{2}r^{3}d\phi\symbol{94}d\tau/\gamma^{6}\\
-4mdr\symbol{94}d\tau/\delta\gamma & 2rd\theta\symbol{94}d\tau & -2\sin
^{2}(\theta)d\phi\symbol{94}d\tau & 0
\end{array}
\right]
\end{align}

\end{center}

By the Strong Equivalence Principle,%

\begin{align}
&  :\{d[\Gamma]+[\Gamma]\symbol{94}[\Gamma]\}\ \ \ \ +\{\left[  \mathbb{T}%
\right]  \symbol{94}\left[  \Gamma\right]  +\left[  \Gamma\right]
\symbol{94}\left[  \mathbb{T}\right]  \}\ \ \ +\{d[\mathbb{T}]+[\mathbb{T}%
]\symbol{94}\mathbb{[T}]\}\\
&  =\ \ \ \ \ \{\left[  \mathbf{\Phi}_{\Gamma}\right]
\}\ \ \ \ \ \ +\{\left[  Interaction\ 2-forms\right]  \}+\ \ \ \{\left[
\mathbf{\Phi}_{\mathbf{T}}\right]  \}\Rightarrow0,
\end{align}
as well as the other formulas of the general theory of the physical vacuum can
be checked using Maple programs which can be found at

\begin{center}
http://www22.pair.com/csdc/pdf/mapleEP1-isotropicSchwartz.pdf,

http://www22.pair.com/csdc/pdf/mapleEP1b-nonsotropicSchwartz.pdf

and will be published on a CD rom, \cite{vol6}.
\end{center}

\subsection{Summary Remarks}

The idea that has been exploited is that the arbitrary Basis Frame (a linear
form), without metric, can be perturbed algebraically to produce a new Basis
Frame that absorbs the properties of a quadratic congruent metric system.
\ This result establishes a constructive existence proof that compatible
metric features of a Physical Vacuum can be derived from the structural format
of the Basis Frame. \ The Basis Frame is the starting point and the congruent
metric properties are deduced.

For the Schwarzschild example, another remarkable feature is that the 1-forms
$\left\vert \sigma^{k}\right\rangle $ constructed according to the formula
\begin{equation}
\lbrack\widehat{\mathbb{B}}]\circ\left\vert dy^{a}\right\rangle \Rightarrow
\left\vert \sigma^{k}\right\rangle =\left\vert A^{k}\right\rangle ,
\end{equation}
are all integrable (as the Topological Torsion term is Zero), but the
coefficients of affine torsion are not zero. \ The symbol $\left\vert
dy^{a}\right\rangle $ stands for the set $[dr,d\theta,d\varphi,d\tau]$
(transposed into a column vector), and $[\widehat{\mathbb{B}}]$ is the
"perturbed" Basis Frame which contains the Schwarzschild metric as a congruent
symmetry. \ The integrability condition means that there exist integrating
factors $\lambda^{(k)}$ for each $\sigma^{k}$ such that a new Basis Frame can
be constructed from\ $[\widehat{\mathbb{B}}]$ algebraically. \ Relative to
this new Basis Frame, the vector of torsion 2-forms is zero, $\left\vert
d\sigma^{k}\right\rangle =\left\vert dA^{k}\right\rangle =\left\vert
F^{k}\right\rangle =0$! \ The "Coriolis" acceleration which is related to the
2-form of torsion 2-forms $\left\vert F^{k}\right\rangle $ can be eliminated
algebraically.! \ Although this result is possible algebraically, it is not
possible diffeomorphically. \ 

\ Of course, this algebraic reduction is impossible if any of the 1-forms,
$\sigma^{k}$, is of Pfaff dimension 3 or more. \ The Basis Frame then admits
Topological Torsion, which is irreducible. \ 

\section{Example 2. \ [$\mathbb{B}$] as a 13 parameter group}

\subsection{\textbf{The Intransitive "Wave -Affine" Connection}}

The next set of examples considers the structure of those 4 x 4 Basis Frames
that admit a 13 parameter group in \ 4 geometrical dimensions of space-time.
\ There are 3 interesting types of 13 parameter group structures. \ This first
example utilizes the canonical form of the 13 parameter "Wave Affine" Basis
Frame. \ These Basis Frames will have zeros for the first 3 elements of the
right-most column. \ Wave Affine Basis Frames exhibit closure relative to
matrix multiplication. \ All products of Wave Affine Basis Frames have 3 zeros
on the right column. \ From a projective point of view, these matrices are not
elements of a transitive group. \ They are intransitive and have fixed points.
\ The true affine group in 4 dimensions is a transitive group (without fixed
points)\ and is discussed in the next subsection..

For simplicity in display, the 9 parameter space-space portions of the Basis
Frame will be assumed to be the $3\times3$ Identity matrix, essentially
ignoring spatial deformations and spatially extended rigid body motions. \ In
the language of projective geometry, this intransitive system has a fixed
point. \ The first 3 elements in the bottom row can be identified (formally)
with the components of a vector potential in electromagnetic theory. The 4th
(space-time) column will have three zeros, and the $\mathbb{B}_{4}^{4}$
component will be described in terms of a function $\phi(x,y,z,t).$%

\begin{equation}
\text{ }\left[  \mathbb{B}_{wave\_affine}\right]  =\left[
\begin{array}
[c]{cccc}%
1 & 0 & 0 & 0\\
0 & 1 & 0 & 0\\
0 & 0 & 1 & 0\\
Ax & Ay & Az & -\phi
\end{array}
\right]  .
\end{equation}
The projected 1-forms become%

\begin{equation}
\lbrack\mathbb{B}_{wave\_affine}]\circ\left\vert dy^{a}\right\rangle
=\left\vert A^{k}\right\rangle
\end{equation}%
\begin{align}
\left[
\begin{array}
[c]{cccc}%
1 & 0 & 0 & 0\\
0 & 1 & 0 & 0\\
0 & 0 & 1 & 0\\
A_{x} & A_{y} & A_{z} & -\phi
\end{array}
\right]  \circ\left\vert
\begin{array}
[c]{c}%
dx\\
dy\\
dz\\
dt
\end{array}
\right\rangle  &  \Rightarrow\left\vert
\begin{array}
[c]{c}%
dx\\
dy\\
dz\\
Action
\end{array}
\right\rangle =\left\vert A^{k}\right\rangle ,\\
Action  &  =A_{x}dx+A_{y}dy+A_{z}dz-\phi dt
\end{align}
The exterior derivative of the vector of 1-forms, $\left\vert A^{k}%
\right\rangle $, produces the vector of 2-forms representing the field
intensities $\left\vert F^{k}\right\rangle $ of electromagnetic theory,
\begin{equation}
\text{\textbf{Intensity 2-forms:} \ }d\left\vert A^{k}\right\rangle
=\left\vert F^{k}\right\rangle =\left\vert
\begin{array}
[c]{c}%
dx\\
dy\\
dz\\
d(Action)
\end{array}
\right\rangle .
\end{equation}
Note that the Action 1-form produced by the wave affine Basis Frame is
precisely the format of the 1-form of Action used to construct the
Electromagnetic field intensities in classical EM theory.

The Cartan right Connection matrix of 1-forms based upon the Basis Frame,
$\left[  \mathbb{B}_{wave\_affine}\right]  ,$ given above is given by the expression%

\begin{equation}
\left[  \mathbb{C}_{wave\_affine\_right}\right]  =\left[
\begin{array}
[c]{cccc}%
0 & 0 & 0 & 0\\
0 & 0 & 0 & 0\\
0 & 0 & 0 & 0\\
-d(A_{x})/\phi & -d(A_{y})/\phi & -d(A_{z})/\phi & d(\ln\phi)
\end{array}
\right]  .
\end{equation}
The general theory permits the vector of excitation 2-forms, $\left\vert
G\right\rangle $, to be evaluated as:%
\begin{align}
\left\vert \Sigma_{W-Affine\_torsion}\right\rangle  &  =\left[  \mathbb{C}%
_{wave\_affine\_right}\right]  \symbol{94}\left\vert dy^{m}\right\rangle
\simeq\left\vert G\right\rangle \\
\text{ \textbf{Excitation 2-forms} }  &  \left\vert G\right\rangle
=\left\vert
\begin{array}
[c]{c}%
0\\
0\\
0\\
-(F)/\phi
\end{array}
\right\rangle =\left\vert
\begin{array}
[c]{c}%
0\\
0\\
0\\
-d(Action)/\phi
\end{array}
\right\rangle \\
&  \neq\left\vert \Sigma_{Car\tan\_torsion}\right\rangle
\end{align}
The coefficients of the vector of excitation 2-forms are precisely those
ascribed to the coefficients of "Affine Torsion", even though the Basis Frame
used as the example is not a member of the transitive affine group.

The exterior derivative of the vector of two forms $\left\vert G\right\rangle
$ produces the vector of 3-form currents $\left\vert J\right\rangle $. \ The
result is%

\begin{equation}
\left\vert J\right\rangle =\left\vert
\begin{array}
[c]{c}%
0\\
0\\
0\\
d(-F)/\phi
\end{array}
\right\rangle =\left\vert
\begin{array}
[c]{c}%
0\\
0\\
0\\
+d\phi\symbol{94}F/\phi^{2}%
\end{array}
\right\rangle
\end{equation}
Note that for this example, the current 3-form is closed producing a
conservation law:%

\begin{equation}
d\left\vert J\right\rangle =\left\vert 0\right\rangle
\end{equation}

The \textbf{wave} \textbf{affine group} of Basis Frames supports a vector of
"Affine Torsion" 2-forms that are abstractly related to excitation 2-forms
$\left\vert G\right\rangle $ representing the fields (the $\mathbf{D}$ and
$\mathbf{H}$ fields) generated by the sources in classical EM theory. \ The
charge-current 3-form need not vanish. \ The 3-forms of Topological Spin and
Topological Torsion are proportional to one another for the simple example,
and are not zero if the Pfaff Topological Dimension of the 1-form of Action is
3 or more. \ Such systems are not in thermodynamic equilibrium.%
\begin{align*}
\text{Topological Torsion}  &  \text{: \ }A\symbol{94}F\\
\text{Topological Spin}  &  \text{: \ }A\symbol{94}G=-A\symbol{94}F/\phi
\end{align*}
If the Poincare invariants are to vanish it is necessary that the 4-form of
Topological Parity is zero, $F\symbol{94}F\Rightarrow0.$

These results for the 13 parameter intransitive "Wave Affine group" are to be
compared to the results for the 13 parameter transitive "Particle Affine
group" given in the next example. \ Maple programs for both types are available.

\subsection{\textbf{The Transitive "Particle -Affine" Connection}}

The next example utilizes the canonical form of the 13 parameter "Particle
Affine" Basis Frame, which will have zeros for the first 3 elements of the
fourth row.\ For simplicity, the 9 parameter space-space portions of the Basis
Frame will be assumed to be the $3\times3$ Identity matrix,
essentially\ ignoring spatial deformations and spatial\ extended rigid body
motions. \ In the language of projective geometry, the transitive system has
no fixed point. \ The 4th (space-time) column will consist of components that
can be identified (formally) with a velocity field. \ The products of
particle-affine matrices exhibit structural closure relative to matrix
multiplication. \ The bottom row always will have three zeros for the first
three matrix elements, and the $\mathbb{B}_{4}^{4}$ component will be
described in terms of a function $\psi(x,y,z,t).$%

\begin{equation}
\text{ }\left[  \mathbb{B}_{Particle\_affine}\right]  =\left[
\begin{array}
[c]{cccc}%
1 & 0 & 0 & -V^{x}\\
0 & 1 & 0 & -V^{y}\\
0 & 0 & 1 & -V^{z}\\
0 & 0 & 0 & \psi
\end{array}
\right]  .
\end{equation}
The projected 1-forms become%

\begin{equation}
\lbrack\mathbb{B}]\circ\left\vert dy^{a}\right\rangle =\left\vert \sigma
^{k}\right\rangle
\end{equation}%
\begin{align}
\left[
\begin{array}
[c]{cccc}%
1 & 0 & 0 & -V^{x}\\
0 & 1 & 0 & -V^{y}\\
0 & 0 & 1 & -V^{z}\\
0 & 0 & 0 & \psi
\end{array}
\right]  \circ\left\vert
\begin{array}
[c]{c}%
dx\\
dy\\
dz\\
dt
\end{array}
\right\rangle  &  \Rightarrow\left\vert
\begin{array}
[c]{c}%
\sigma^{x}\\
\sigma^{y}\\
\sigma^{z}\\
\omega
\end{array}
\right\rangle \\
&  =\left\vert
\begin{array}
[c]{c}%
dx-V^{x}dt\\
dy-V^{y}dt\\
dz-V^{z}dt\\
\psi dt
\end{array}
\right\rangle =\left\vert
\begin{array}
[c]{c}%
\Delta x\\
\Delta y\\
\Delta z\\
\psi dt
\end{array}
\right\rangle ,
\end{align}
where the terms $\Delta x,~\Delta y,\Delta z$, if not zero, can be considered
as the topological fluctuations in "kinematic perfection". \ \ The exterior
derivative of the vector of 1-forms, $\left\vert A^{k}\right\rangle $,
produces the vector of 2-forms representing the field intensities $\left\vert
F^{k}\right\rangle $ of electromagnetic theory,
\begin{equation}
\text{Intensity 2-forms: \ }d\left\vert A^{k}\right\rangle =\left\vert
F^{k}\right\rangle =\left\vert
\begin{array}
[c]{c}%
-dV^{x}\symbol{94}dt\\
-dV^{y}\symbol{94}dt\\
-dV^{z}\symbol{94}dt\\
d\psi\symbol{94}dt
\end{array}
\right\rangle =\left\vert
\begin{array}
[c]{c}%
d(\Delta x)\\
d(\Delta y)\\
d(\Delta z)\\
d\psi\symbol{94}dt
\end{array}
\right\rangle .
\end{equation}
If the fluctuations about kinematic perfection are closed, or the velocity
field is a function of the single parameter of time, and if the function
$\psi$ is not dependent upon the spatial variables, then the vector of field
intensities is zero; \ \textbf{E} and \textbf{B} fields do not exist in this
refined topology of kinematic perfection and universal time, based upon the
transitive affine group. \ If the potential function $\psi$ is not spatially
uniform, then there can exist the equivalent of an Electric Field, but there
never appear Magnetic fields (for the transitive affine Basis Frame without
spatial deformations).

The Cartan right Connection matrix of 1-forms for this refined Basis Frame is
given by the expression%

\begin{equation}
\left[  \mathbb{C}_{Particle\_affine\_right}\right]  =\left[
\begin{array}
[c]{cccc}%
0 & 0 & 0 & -dV^{x}+V^{x}d(\ln\psi)\\
0 & 0 & 0 & -dV^{y}+V^{y}d(\ln\psi)\\
0 & 0 & 0 & -dV^{z}+V^{z}d(\ln\psi)\\
0 & 0 & 0 & d(\ln\psi)
\end{array}
\right]
\end{equation}

The Connection coefficients can be computed and exhibit components of non zero
"P-Affine Torsion ", especially if spatial deformations exist, or the
potential, $\psi,$ is spatially dependent. \ The vector of "Affine Torsion"
2-forms is given by the expressions:%

\begin{align}
\left\vert \Sigma_{P-Affine\_torsion}\right\rangle  &  =\mathbb{C}%
\symbol{94}\left\vert dy^{m}\right\rangle \simeq\left\vert G_{Particel}%
\right\rangle \\
\text{\textbf{\ field excitations} }  &  \text{:}\left\vert G_{Particel}%
\right\rangle =\left\vert
\begin{array}
[c]{c}%
-d(V^{x})\symbol{94}d(t)+V^{x}d(\ln\psi)\symbol{94}d(t))\\
-d(V^{y})\symbol{94}d(t)+V^{y}d(\ln\psi)\symbol{94}d(t))\\
-d(V^{z})\symbol{94}d(t)+V^{z}d(\ln\psi)\symbol{94}d(t))\\
d(\ln\psi)\symbol{94}d(t)
\end{array}
\right\rangle \\
&  \neq\left\vert \Sigma_{\text{Cartan\_torsion}}\right\rangle
\end{align}
If the velocity field is a function of time only, $V=V(t)$, then total
differential of the velocity field leads to the classic kinematic concept of
accelerations, and the "affine" torsion 2-forms depend only on the potential
$\psi.$ \ If the potential function $\psi$ is such that its total differential
is zero, or a function of time (and not dependent on the spatial coordinates),
then all of the affine torsion coefficients vanish (for this example).
\ Moreover, under the kinematic assumption, the 3-forms of currents vanish:%

\begin{equation}
\left\vert J\right\rangle =d\left\vert G_{Particel}\right\rangle \Rightarrow0.
\end{equation}

Now consider the case of topological fluctuations, $\Delta x^{k},$ about
kinematic perfection. \ That is suppose%

\begin{align}
dx^{k}-V^{k}dt  &  =\Delta x^{k},\\
\text{or \ }\left\vert
\begin{array}
[c]{c}%
\mathbf{\sigma}^{k}\\
\omega
\end{array}
\right\rangle  &  =\left\vert
\begin{array}
[c]{c}%
\Delta\mathbf{x}^{k}\\
\psi dt
\end{array}
\right\rangle \\
\text{such that \ }\left\vert
\begin{array}
[c]{c}%
d\mathbf{\sigma}^{k}\\
d\omega
\end{array}
\right\rangle  &  =\left\vert
\begin{array}
[c]{c}%
d(\Delta\mathbf{x}^{k})\\
d\psi\symbol{94}dt
\end{array}
\right\rangle .
\end{align}
The algebra of this example based on topological fluctuations about kinematic
perfection are discussed in the Maple programs which can be found at

\begin{center}
http://www22.pair.com/csdc/pdf/mapleEP2-waveaffine.pdf,

http://www22.pair.com/csdc/pdf/mapleEP2-particleaffine.pdf

and will be published on a CD rom, \cite{vol6}.
\end{center}

The Maple programs construct the more tedious algebraic results (that are
related to curvatures and the Bianchi identities) developed in Section 2 above.

\ The \textbf{particle affine} group of Basis Frames has a connection that is
compatible with the concept of kinematic evolution of massive particles.

\section{Example 3. \ [$\mathbb{B}$] as an element of the Lorentz Group}

In the theory of Electromagnetic Signals, it was an objective of Fock
\cite{Fock} to establish the equivalence class of coordinates, or
diffeomorphic systems of reference, where by if one observer in one system of
reference claims to "see" an electromagnetic signal (defined as a propagating
discontinuity), then another observer in a different reference system would
make a similar statement with regard to the same physical discontinuity
phenomenon. \ Both observers claim to see a "signal"; \ they both see a
propagating discontinuity in field strength.\ \ The propagating discontinuity
is defined by the solution to a non-linear first order partial differential
equation, known as the Null Eikonal equation \cite{vol4} that describes the
point set upon which the solutions to the Maxwell system of PDE's are not
unique. \ The Null Eikonal equation is a quadratic sum of partial
differentials with the Minkowski signature:
\begin{equation}
\text{\textbf{Null Eikonal}}\ \ \ (\pm\partial\varphi/\partial x)^{2}%
\pm(\partial\varphi/\partial y)^{2}\pm(\partial\varphi/\partial z)^{2}%
\mp1/c^{2}(\partial\varphi/\partial t)^{2}=0.
\end{equation}
To preserve the discontinuity, the Null Eikonal equation must be preserved
under the diffeomorphisms that relate one observer to another. \ 

The Null Eikonal equation can be formulated as a congruent product of a vector
of 1-forms. \ %

\begin{align}
\left\vert (d\varphi)^{k}\right\rangle  &  =\left[
\begin{array}
[c]{cccc}%
(\partial\varphi/\partial x) & 0 & 0 & 0\\
0 & (\partial\varphi/\partial y) & 0 & 0\\
0 & 0 & (\partial\varphi/\partial z) & 0\\
0 & 0 & 0 & (\partial\varphi/c\partial t)
\end{array}
\right]  \circ\left\vert
\begin{array}
[c]{c}%
dx\\
dy\\
dz\\
dt
\end{array}
\right\rangle \\
\text{Null Eikonal Equation }  &  \text{:}0=\left\langle d\varphi\right\vert
\circ\left[  \eta\right]  \circ\left\vert d\varphi\right\rangle \Rightarrow
\left\langle d\varphi\right\vert \circ\left[  \mathbb{L}\right]  ^{T}%
\circ\left[  \eta\right]  \circ\left[  \mathbb{L}\right]  \circ\left\vert
d\varphi\right\rangle =0.
\end{align}
If the infinitesimal vector of 1-forms $\left\vert (d\varphi)^{k}\right\rangle
$ is transformed to a new set of 1-forms by the Basis Matrix $\left[
\mathbb{L}\right]  ,$ it is apparent that the Null Eikonal equation is
invariant in form, when $\left[  \mathbb{L}\right]  $ is a Lorentz
transformation:
\begin{equation}
\text{Lorentz automorphism: \ }\left[  \mathbb{L}\right]  ^{T}\circ\left[
\eta\right]  \circ\left[  \mathbb{L}\right]  =\left[  \eta\right]  .
\end{equation}

\ Fock established that the only linear transformation that would preserve
this equivalence was the Lorentz transformation. \ However, there is a larger
non-linear class of transformations that also preserves the propagating
discontinuity as a discontinuity. \ \ The key requirement such that the
propagating field amplitude \textit{discontinuity} is preserved is that the
null line element (the zero set) must be preserved. \ It is not the quadratic
metric form that must be preserved, it is only the zero set of the quadratic
form that must be preserved. \ The zero set of the Null Eikonal equation is
also satisfied relative to the conformal extensions of the Lorentz transformation.:%

\begin{align}
\text{\textbf{Extended }}  &  :\text{\textbf{Lorentz transformation} }\\
\left[  \lambda\mathbb{L}\right]  ^{T}\circ\left[  \eta\right]  \circ\left[
\lambda\mathbb{L}\right]   &  =\lambda^{2}\left[  \eta\right]  \text{,}\\
\left\langle d\varphi\right\vert \circ\left[  \eta\right]  \circ\left\vert
d\varphi\right\rangle  &  \Rightarrow0\supset\lambda^{2}\left\langle
d\varphi\right\vert \circ\left[  \eta\right]  \circ\left\vert d\varphi
\right\rangle \Rightarrow0.
\end{align}
Fock demonstrated that the extended Conformal (non linear) Lorentz
transformations also preserve the the propagating electromagnetic
discontinuity. \ In this example of the "Physical Vacuum" conformal non-linear
Lorentz transformations will be studied as a class of Basis Frames suitable
for description of a specialized "Physical Vacuum". \ The surprise is that the
method leads to a better understanding not only of the theory of
electromagnetic signals as propagating amplitude\ discontinuities, but also to
the theory of propagating tangential discontinuities such as Wakes, in a field
continuum, or a fluid. \ 

\subsubsection{Harmonic Wakes}

From the topological point of view it is remarkable how often flow
instabilities and wakes take on one or another of two basic scroll patterns.
The first scroll pattern is epitomized by the Kelvin-Helmholtz instability
(Figure 1a) and the second scroll pattern is epitomized by the Raleigh-Taylor
instability (Figure 1b)

\begin{center}
\bigskip%
{\includegraphics[
height=1.4347in,
width=3.039in
]%
{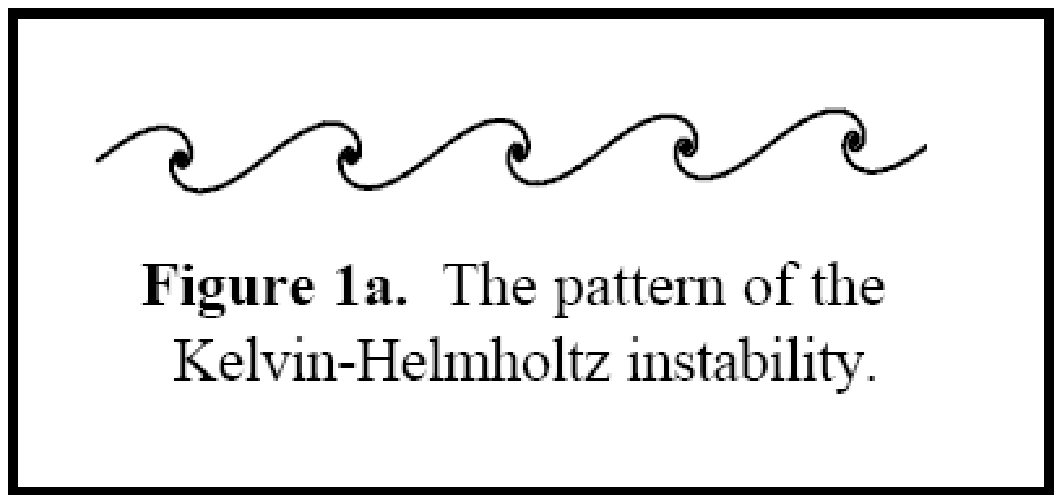}%
}%
%

{\includegraphics[
height=1.4209in,
width=3.09in
]%
{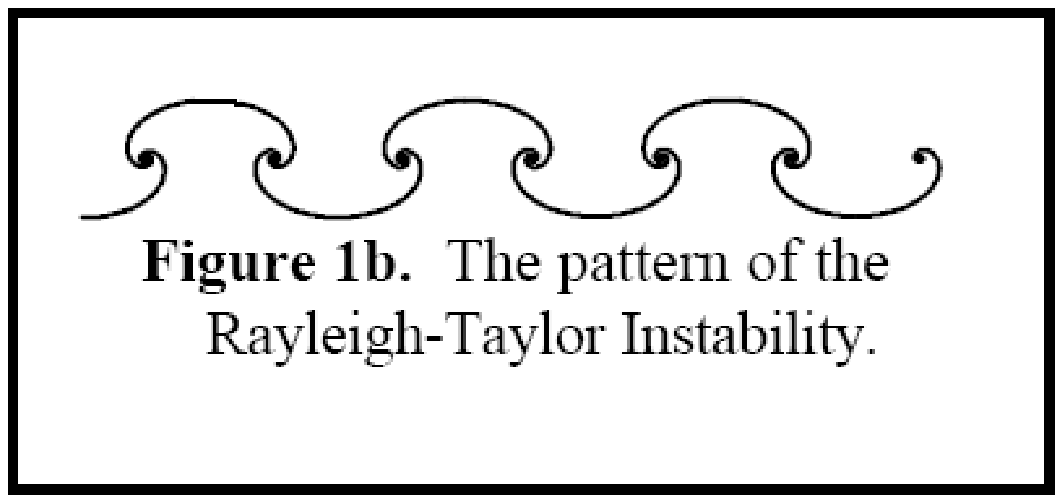}%
}%

\end{center}

The repeated occurrence of these two scroll patterns (one similar to a
replication of a Cornu Spiral and the other similar to a replication of a
Mushroom Spiral), often deformed but still recognizable and persistent, even
in a dissipative environment, suggests that a basic simple underlying
topological principle is responsible for their creation. \ The essential
questions are: Why do these spiral patterns appear almost universally in
wakes? Why do they persist for such substantial periods of time? Why are they
so sharply defined? What are the details of their creation? As H.K.Browand has
said (p. 117 in \ \cite{Browand}) "There does not exist a satisfactory
theoretical explanation for these wake patterns." \ 

The mushroom pattern is of particular interest to this author, who long ago
was fascinated by the long lived ionized ring that persists in the mushroom
cloud of an atomic explosion. Although the mushroom pattern appears in many
diverse physical systems (in the Frank-Reed source of crystal growth, in the
scroll patterns generated in excitable systems, in the generation of the wake
behind an aircraft,...), no simple functional description of the deformable
mushroom pattern as a topological property was known to me in 1957. \ My first
exposure to clothoids was about\ 1995 \cite{Gray}. \ Classical geometric
analysis applied to the equations of hydrodynamics failed to give a
satisfactory description of these persistent structures, so often observed in
many different situations. \ It is remarkable that Frenet theory concepts not
only yields a quantitative picture of the odd and even scroll patterns, but
also gives closed form solutions for the creation of the limit set figures above.

An argument\ made in (1985-1990) for the existence of long lived wakes was
extracted from fluids described phenomenologically in terms of the
Navier-Stokes equations. \ For incompressible fluid dynamics, with
$div\mathbf{V}=0$, an initial (perhaps turbulent) fluid velocity distribution
would decay by shear viscosity processes, such as those encoded by the
Navier-Stokes term, $\nu\nabla^{2}\mathbf{V}\neq0.$ \ However, any components
of the initial velocity field that are harmonic, $\nabla^{2}\mathbf{V}=0,$
will not decay, and it was argued that these components of the initial
velocity distribution are those that form a residue or wake. \ The residue
velocity fields admit a Hamiltonian representation, and therefor have a
persistent existence. \ The important fundamental observation of differential
geometry was that a harmonic vector field describes a minimal surface. \ Hence
wakes should be related (somehow) to flows that describe minimal surfaces in
velocity space. \ The first analytic example is discussed in Chapter 8.2.4 of
\cite{vol3}.

The observable features of hydrodynamic wakes can be put into correspondence
with those characteristic surfaces of tangential discontinuities upon which
the solutions to the evolutionary equations of hydrodynamics are not unique.
Only the robust minimal surface subset, associated with a harmonic vector
field, will be persistent and of minimal dissipation. Surprisingly, those
minimal surfaces if generated by iterates of complex holomorphic curves in
four dimensions are related to fractal sets.

The clue that wave fronts, representing propagating limit sets of tangential
discontinuities, are the basis for the two basic spiral (or scroll)
instability patterns in fluid dynamics came to this author during a study of
Cartan's methods of differential topology and the Frenet-Cartan concept of the
Repere Mobile as applied to the production of defects and topological torsion
in dynamical systems \cite{rmkmoffat} \cite{rmkpermb} \cite{RMKsectam}.

At first consider Frenet space curves in the plane, which are generated by a
pair of differential equations of the form,%

\begin{align}
dx/ds  &  =\text{sin(}Q(s)),\ \label{wa2}\\
\ \ \ \ \ \ \ \ dy/ds  &  =\text{cos}(Q(s)), \label{wa3}%
\end{align}
The RHS of these equations is always a unit vector, which encourages the
identification of this formalism with the Frenet formulas for motion in a plane.%

\begin{equation}
d\mathbf{R}/ds=\left\vert
\begin{array}
[c]{c}%
dx/ds\\
dy/ds
\end{array}
\right\rangle =\mathbf{t}(s)=\left\vert
\begin{array}
[c]{c}%
sin(Q(s))\\
cos(Q(s))
\end{array}
\right\rangle \
\end{equation}
By constructing the differential of the unit vector, $\mathbf{t}(s)\mathbf{,}
$\ with respect to the "arclength" $s$ leads to the classic expression for the
Frenet curvature:
\begin{align}
d\mathbf{t}(s)/ds  &  =\kappa\mathbf{n}(s)\mathbf{=\{}dQ/ds\}\left\vert
\begin{array}
[c]{c}%
\text{cos}(Q(s))\\
-\text{sin}(Q(s))
\end{array}
\right\rangle ,\ \\
\kappa &  =\mathbf{\{}dQ/ds\}.
\end{align}
For $\kappa=1,$ integration yields the space curve which is a circle. \ For
$\kappa=1/s$, the resulting space curve is the logarithmic spiral in the x-y
plane. For $\kappa=s$, the resulting image in the x-y plane is the Cornu
spiral. These facts have been known for more than 100 years to differential geometers.

However, a simple sequence is to be recognized :
\begin{align}
\ \ \  &  :\ \ \kappa=s^{-1},\ \ \ \ \ \ \ \ \ \ \kappa=s^{0}%
,\ \ \ \ \ \ \ \ \kappa=s^{1}...\\
&  :\text{Log spiral, \ \ \ \ \ \ \ circle, \ \ \ \ \ \ Cornu-Fresnel
spiral...}\nonumber
\end{align}
An extension of the sequence raises the question: what are the characteristic
shapes of intrinsic space curves for which the curvature is proportional to an
arbitrary power of the arc length, $\kappa=g(s)=s^{n}$, for all positive and
negative integers (or rational fractions)? \ Through the power of the PC these
questions may be answered quickly by integrating the Harmonic Ordinary
Differential equations.

Suppose the argument function is given by the expression,%

\begin{align}
Q  &  =s^{n+1}/(n+1),\text{ such that}\\
dQ/ds  &  =\kappa=s^{n}.
\end{align}
The results of the numerical integrations are presented in Figure 2 for n = 1
and n = 2. A most remarkable result is that the Cornu-Fresnel spiral of Figure
2a is the deformable equivalent for all odd-integer n \texttt{%
$>$%
} 0, and the Mushroom spiral of Figure 2b is the deformable equivalent for all
even-integer n
$>$
0.

\begin{center}
\bigskip%
{\includegraphics[
height=2.2667in,
width=3.0165in
]%
{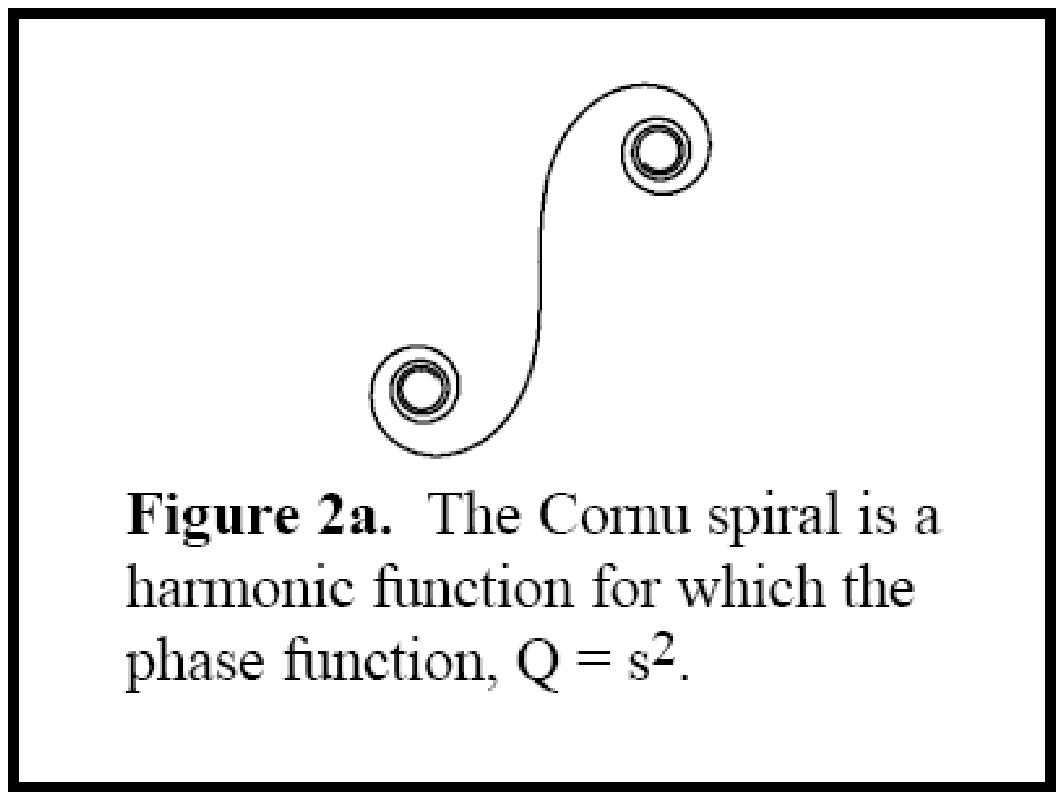}%
}%
%

{\includegraphics[
height=2.2667in,
width=3.0467in
]%
{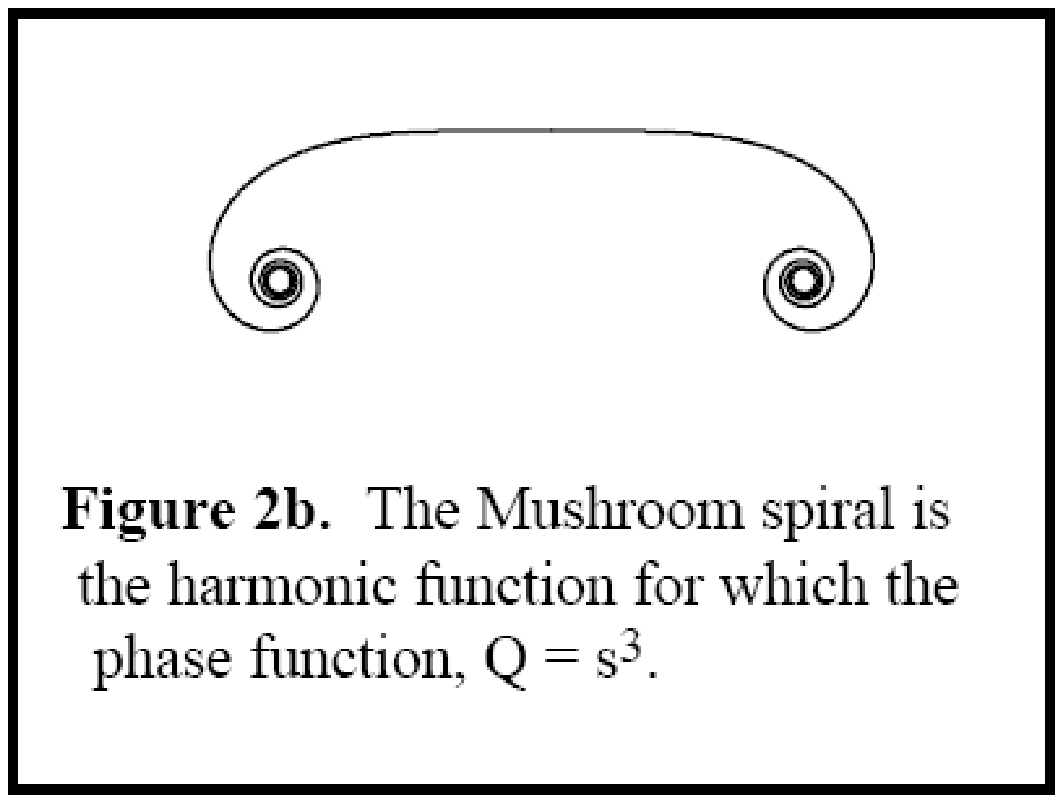}%
}%

\end{center}

Not only has the missing analytic description of the mushroom spiral been
found, but also a \textit{raison d'\^{e}tre} has been established for the
universality of the two spiral patterns. They belong to the even and odd
classes of arc length exponents describing plane curves in terms of the
formula, $\kappa=s^{n}$. The Cornu-Fresnel spiral is the first odd harmonic
function that maps the infinite interval into a bounded region of the plane,
and the mushroom spiral is the first even function that maps the infinite
interval into a bounded region of the plane. Note that the zeroth harmonic
function maps the infinite interval into the bounded region of the plane, but
the trajectory is unique in that it is not only bounded but it is closed;
i.e., a circle. A similar sequence can be generated for the half-integer
exponents with n = +3/2, 7/2, 11/2... giving the Mushroom spirals and
5/2,9/2,13/2... giving the Cornu spirals. In numeric and experimental studies
of certain shear flows, the Cornu spiral appears to dominate the motion in the
longitudinal direction, while the Mushroom spiral appears in the transverse direction.

\begin{remark}
\textit{It is important to remember that the Cornu spirals are related to
diffraction of waves.}
\end{remark}

Periodic patterns can be obtained by examining phase functions of various
forms. For example, the Kelvin Helmholtz instability of Figure 1a is
homeomorphic to the choice Q(s) = 1/cos$^{2}$(s). \ Similarly, the
Rayleigh-Taylor instability of Figure 1b is homeomorphic to the case Q(s) =
tan(s)/cos(s). \ 

\begin{center}%
{\includegraphics[
height=1.5437in,
width=3.0234in
]%
{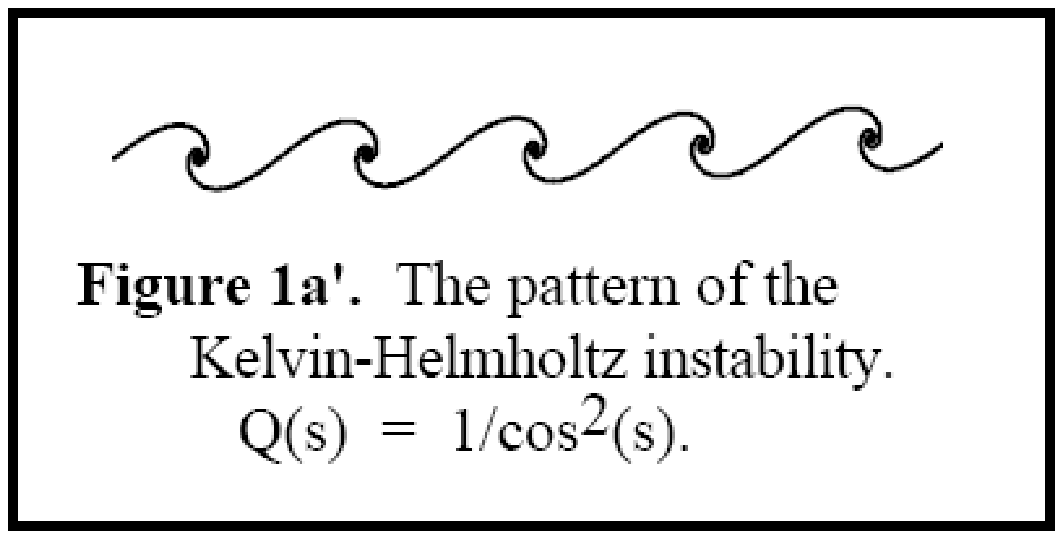}%
}%
%

{\includegraphics[
height=1.5515in,
width=3.0165in
]%
{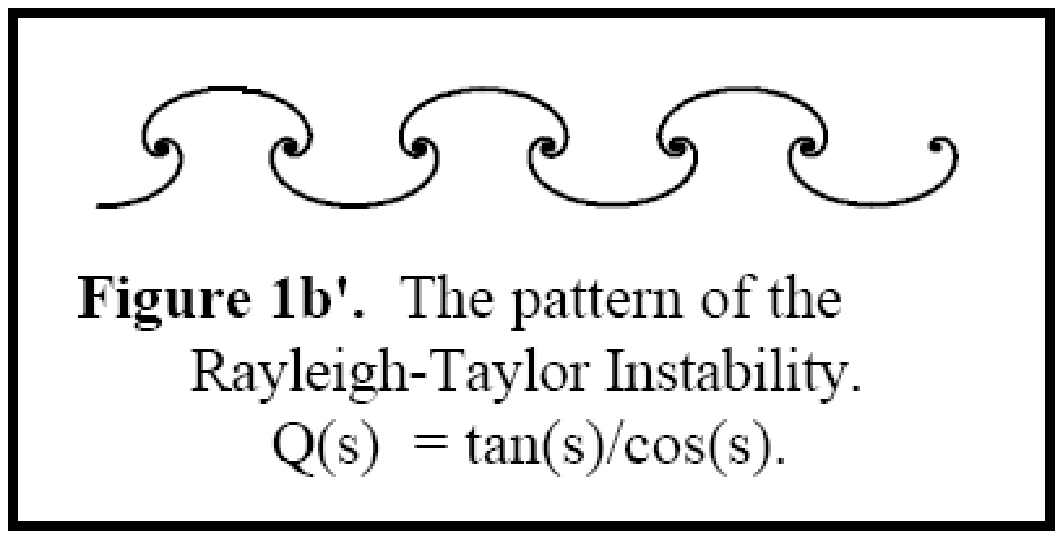}%
}%

\end{center}

These shapes were found by numerical integration (on a 100Mhz PC) of the
Frenet space curve equations in a plane. \ The integration technique made use
of the arclength $s$ as the fundamental integration parameter. \ 

\begin{remark}
What has all this to do with Conformal Lorentz transformations? \ As will be
developed below, the harmonic forms Q(s) generating the harmonic wake patterns
also appear as the canonical matrix elements in extended Lorentz transformations!
\end{remark}

\subsection{Conformal Lorentz Transformations}

The problem is reduced to the study of those transformations $\left[
\mathbb{L}\right]  $ acting on the differential position vectors $\left\vert
dx^{a}\right\rangle $ that preserve a certain quadratic form (the square of
the infinitesimal arclength) as an invariant of the transformation, to within
a factor. \ By definition of the Physical Vacuum, a Basis Frame $\left[
\mathbb{B}\right]  $ acting on a vector array of perfect differentials
$\left\vert dx^{a}\right\rangle $ will produce a vector array of 1-forms,
$\left\vert \sigma^{k}\right\rangle .$ \ These 1-forms need not be exact, and
not even closed. \
\begin{equation}
\left[  \mathbb{B}_{a}^{k}\right]  \circ\left\vert dx^{a}\right\rangle
=\left[  \mathbb{B}\right]  \circ\left\vert
\begin{array}
[c]{l}%
dx\\
dy\\
dz\\
cdt
\end{array}
\right\rangle \Rightarrow\left\vert \sigma^{k}\right\rangle .
\end{equation}

The concept of an infinitesimal arclength squared can be defined in terms of a
congruent quadratic measure defined with respect to a diagonal Sylvestor
signature matrix: \ \
\begin{align}
(\delta s)^{2}  &  =\left\langle dx^{a}\right\vert \circ\left[  \mathbb{B}%
\right]  ^{Transpose}\circ\left[  \eta\right]  \circ\left[  \mathbb{B}\right]
\circ\left\vert dx^{a}\right\rangle \\
&  =\left\langle dx^{a}\right\vert \circ\left[  g\right]  \circ\left\vert
dx^{a}\right\rangle =\left\langle \sigma^{k}\right\vert \circ\left[
\eta\right]  \circ\left\vert \sigma^{k}\right\rangle \\
&  =\mathbf{B}_{m}^{1}dx^{m}\eta_{11}\mathbb{B}_{n}^{1}dx^{n}+\mathbf{B}%
_{m}^{2}dx^{m}\eta_{22}\mathbb{B}_{n}^{2}dx^{n}....
\end{align}
The infinitesimal arclength, $\delta s$, can be exact, not closed, or even a
non-integrable 1-form. A null length is given by the expression, $(\delta
s)^{2}=0$. \ 

Now restrict the group of Basis Frames $\left[  \mathbb{B}\right]  $ to be
those (to within a factor) that belong to the Lorentz group $\left[
\mathbb{L}\right]  $ such that the quadratic congruence is invariant about the
identity. \ The idea is that
\begin{align}
\left[  \eta\right]   &  \Leftarrow\left[  \mathbb{L}\right]  ^{Transpose}%
\circ\left[  \eta\right]  \circ\left[  \mathbb{L}\right]  =\left[
\eta\right]  ,\\
(\delta s)^{2}  &  =\left\langle dx^{k}\right\vert \circ\left[  \eta\right]
\circ\left\vert dx^{k}\right\rangle =\\
&  =\left\langle dx^{k}\right\vert \circ\left[  \mathbb{L}\right]
^{Transpose}\circ\left[  \eta\right]  \circ\left[  \mathbb{L}\right]
\circ\left\vert dx^{k}\right\rangle =(\delta s^{\prime})^{2}\\
&  =\mp(dx)^{2}\mp(dy)^{2}\mp(dz)^{2}\pm c^{2}(dt)^{2}.
\end{align}
\ The Lorentz transformation is a correlation automorphism with respect to the
Minkowski metric, $\left[  \eta\right]  $. \ 

However, to preserve the zero set of the line element, such that
\begin{equation}
(\delta s^{\prime})^{2}\Rightarrow0\supset(\delta s)^{2}\Rightarrow0,
\end{equation}
\ all that is required is that \vspace{1pt}the transformation\ matrix be an
element of the group of automorphisms (relative to $\left[  \eta\right]  )$ to
within a factor, $\lambda:$
\begin{equation}
\left[  \mathbb{L}\right]  ^{transpose}\circ\left[  \eta\right]  \circ\left[
\mathbb{L}\right]  =\lambda^{2}(x,y,z,t)\left[  \eta\right]  .
\end{equation}
The factor $\lambda$ can be an arbitrary function of the variables $\{x,y,z,t
$\}. \ \ 

The fundamental concept that a signal remains a signal is equivalent to the
idea that the Minkowski metric (or better said, the quadratic form of the
Eikonal equation) is to be preserved, \textit{to within a factor}, by the
congruent mapping presented in the equation above. \ Such mappings,
$\lambda\left[  \mathbb{L}\right]  ,$which need not have constant matrix
elements, are defined herein as \textquotedblright extended\textquotedblright%
\ or non-linear Lorentz transformations. \ The conventional Lorentz
transformations are the restricted subset of the extended Lorentz
transformations, where the constraints require that $\lambda=\pm1\,$and the
matrix array is a set of constants. \ Such restrictions will not be the
general case studied herein; constant matrix elements are considered to be the
exception. \ Extended Lorentz transformations define an equivalence class of
coordinate systems (frames of reference). \ It has been conventional to
describe the restricted class of Lorentz systems as "inertial"\ frames of
reference. \ This concept will be examined (and extended) below.

\subsection{\vspace{1pt}Integrable vs. Non-Integrable Lorentz transformations}

Without exhibiting the functional form of the extended Lorentz
transformations, it is to be observed that the\ vector array of 1-forms,
$\left\vert \sigma^{k}\right\rangle ,$%
\begin{equation}
\left\vert \sigma^{k}\right\rangle =\left[  \mathbb{L}\right]  \circ\left\vert
dx^{k}\right\rangle
\end{equation}
need not be an array of perfect differentials. \ The group requirement is
fixed by the constraint that
\begin{equation}
\left[  \mathbb{L}\right]  ^{transpose}\circ\left[  \eta\right]  \circ\left[
\mathbb{L}\right]  =\lambda^{2}(x,y,z,t)\left[  \eta\right]  .
\end{equation}
\ If the mapping $\left[  \mathbb{L}\right]  $ produces exact differentials,
then the exterior derivative of vector array, $\left\vert \sigma
^{k}\right\rangle ,$ is zero, component by component. \vspace{1pt}
\begin{equation}
\text{Vector array of Closure\ 2-forms}:d\left\vert \sigma^{k}\right\rangle
=(d\left[  \mathbb{L}\right]  )\circ\left\vert dx^{k}\right\rangle
\Rightarrow0.
\end{equation}
If the mapping functions of the matrix $\left[  \mathbb{L}\right]  $ are
constants, then $d\left[  \mathbb{L}\right]  =0,$ and the induced 1-forms
$\left\vert \sigma^{k}\right\rangle $ are always closed. \ Each component of
$\left\vert \sigma^{k}\right\rangle $ then satisfies the Frobenius
integrability condition and each new differential coordinate can be reduced
globally to the exact differential of a single function (case1).

However, the components of the extended transformations $\lambda\left[
\mathbb{L}\right]  )\circ\left\vert dx^{k}\right\rangle $ need not be
constants, and yet the new 1-forms created by the mapping may be integrable
over a large but perhaps limited domain. \ The new 1-forms are closed, but not
exact (case 2), leading to interesting topological situations. \ In other
situations, the components of $\left\vert \sigma^{k}\right\rangle $ may not be
closed, but might admit integrating factors which would make them exact or
closed (case 3). The vector array of closure 2-forms (see the next paragraph
below) need not be zero. \ \ In addition, when the matrix elements of $\left[
\mathbb{L}\right]  $ are not constants, it may be true that the components of
$\left\vert \sigma^{k}\right\rangle $ are not uniquely integrable at all (case
4). \ Each of these four cases should be studied separately. \ Each of the 4
cases defines a different topological set of Lorentz transformations.

From the general theory of the Physical Vacuum, it is to be remarked that if a
subset of the extended Lorentz transformations (with inverse) are used as a
Basis Frame at a point p of $\{x,y,z,t\}$, then it is possible to define a
connection in terms of the right Cartan matrix $\left[  \mathbb{C}\right]  $
of 1-forms created by exterior differentiation of the extended Lorentz matrix
$\left[  \mathbb{L}\right]  \Leftrightarrow\left[  \mathbb{B}\right]  .$
\ \ The induced vector array of closure 2-forms becomes%

\begin{align}
d\left\vert \sigma^{k}\right\rangle  &  =(d\left[  \mathbb{L}\right]
)\circ\left\vert dx^{m}\right\rangle =\left[  \mathbb{L}\right]  \circ\left[
C\right]  \symbol{94}\left\vert dx^{m}\right\rangle \\
&  =L_{b}^{k}\circ\left\vert C_{[mn]}^{b}dx^{m}\symbol{94}dx^{n}\right\rangle
=L_{b}^{k}\circ\left\vert C^{b}\right\rangle
\end{align}
The vector of 2-forms $d\left\vert \sigma^{k}\right\rangle $ are formally
equivalent to the field intensities of electromagnetism $\left\vert
F^{k}\right\rangle .$ \ The vector of 2-forms\ $\left\vert C^{b}\right\rangle
$ is formally equivalent to the field excitations of electromagnetic theory.
\ The coefficients $C_{[mn]}^{b}$ are precisely those used\ in tensor analysis
to define the concept of "Affine Torsion". \ Without "Affine Torsion" there
does not exist electromagnetic field components.

An example of a classic Lorentz transformation (often called a "boost") is
given by the matrix of translational shears,%

\begin{equation}
\left[  \mathbb{L}(\beta)_{xt}\right]  =\left[
\begin{array}
[c]{cccc}%
\frac{1}{\sqrt{1-\beta^{2}}} & 0 & 0 & \frac{\beta}{\sqrt{1-\beta^{2}}}\\
0 & 1 & 0 & 0\\
0 & 0 & 1 & 0\\
\frac{\beta}{\sqrt{1-\beta^{2}}} & 0 & 0 & \frac{1}{\sqrt{1-\beta^{2}}}%
\end{array}
\right]  ,
\end{equation}
where $\beta\,$\ is defined as $\mathbf{v}^{x}/c$, the ratio of the x
component of \textquotedblright velocity\textquotedblright\ and the speed of
light, and is conventionally interpreted as a constant. \ 

As another example of a different Lorentz transformation, consider the matrix
(with unit determinant) representing shears of rotation and/or expansion.%

\begin{equation}
\left[  \mathbb{L}(\theta(s)_{xt}\right]  =\left[
\begin{array}
[c]{cccc}%
\frac{1}{\cos(\theta)} & 0 & 0 & \pm\frac{\sin(\theta)}{\cos(\theta)}\\
0 & 1 & 0 & 0\\
0 & 0 & 1 & 0\\
\pm\frac{\sin(\theta)}{\cos(\theta)} & 0 & 0 & \frac{1}{\cos(\theta)}%
\end{array}
,\right]
\end{equation}
$\allowbreak$Substitute this matrix into the equation defining the Lorentz
automorphism, compute the matrix products to show that, indeed, the matrix
$\left[  \mathbb{L}(\theta(s))_{xt}\right]  $ preserves the Minkowski metric:%

\begin{equation}
\left[  \mathbb{L}(\theta)_{xt}\right]  ^{transpose}\circ\left[  \eta\right]
\circ\left[  \mathbb{L}(\theta)_{xt}\right]  =\left[  \eta\right]  .
\end{equation}

\begin{remark}
Note that the dimensionless function $\theta$ can be an arbitrary function of
all the independent variables,
\begin{equation}
\theta(s)=\theta(x,y,z,t).
\end{equation}

\end{remark}

For example, $\theta(x,y,z,t)$ could be equivalent to the arclength divided by
a constant scale length, $\theta(x,y,z,t)\Rightarrow s/\lambda$. \ 

Use the Lorentz matrix $\left[  L(\theta)_{xt}\right]  $ to compute the differentials:%

\begin{equation}
\left[
\begin{array}
[c]{cccc}%
\frac{1}{\cos\theta} & 0 & 0 & \pm\frac{\sin\theta}{\cos\theta}\\
0 & 1 & 0 & 0\\
0 & 0 & 1 & 0\\
\pm\frac{\sin\theta}{\cos\theta} & 0 & 0 & \frac{1}{\cos\theta}%
\end{array}
\right]  \circ\left\vert
\begin{array}
[c]{l}%
dx\\
dy\\
dz\\
cdt
\end{array}
\right\rangle =\allowbreak\left\vert
\begin{array}
[c]{c}%
\frac{1}{\cos\theta}dx\pm\frac{\sin\theta}{\cos\theta}cdt\\
dy\\
dz\\
\frac{\sin\theta}{\cos\theta}dx\pm\frac{1}{\cos\theta}cdt
\end{array}
\allowbreak\right\rangle =\left\vert \sigma^{k}\right\rangle
\end{equation}
In general, all of the components of $\left\vert \sigma^{k}\right\rangle $ are
not perfect differentials. \ For example\vspace{1pt}, consider
\begin{equation}
\sigma^{1}=\frac{1}{\cos\theta}dx\pm\frac{\sin\theta}{\cos\theta}cdt.
\end{equation}
When the angular function $\theta$ is a constant, then a new coordinate, $X,
$is well defined\ by integration, and its differential yields the expression above.%

\begin{equation}
X=\frac{1}{\cos\theta}x\pm\frac{\sin\theta}{\cos\theta}%
ct,\,\,\,\,\,\,\,\,\,\,\,\sigma^{1}\Rightarrow
dX\,\,\,\,\,\,\,\,\,and\,\ \,\ d\sigma^{1}=0\,
\end{equation}
However, in the general case, the vector of closure 2 forms $d\left\vert
\sigma^{k}\right\rangle $ of field intensities do not vanish:%

\begin{equation}
\text{ Field Intensities \ }d\left\vert \sigma^{k}\right\rangle =\left\vert
\begin{array}
[c]{l}%
-\frac{\sin\theta}{\cos^{2}\theta}dx\symbol{94}d\theta-\frac{1}{\cos^{2}%
\theta}dt\symbol{94}d\theta\\
0\\
0\\
-\frac{\sin\theta}{\cos^{2}\theta}dt\symbol{94}d\theta-\frac{1}{\cos^{2}%
\theta}dx\symbol{94}d\theta
\end{array}
\right\rangle \neq0.
\end{equation}
For the example given, the 4 forms of the type $d\sigma^{k}\symbol{94}%
d\sigma^{k}$ do not vanish (the Topological Torsion is zero as there are only
3 independent differentials, $dx\symbol{94}d\theta\symbol{94}dt$.

Now consider the extended Lorentz transformations given by the matrix%
\begin{align}
\left[  \mathbf{P}(\theta)_{xt}\right]   &  =\lambda\left[  \mathbb{L}%
(\theta)_{xt}\right]  =(1/\cos\theta)\left[  \mathbb{L}(\theta)_{xt}\right] \\
\left[  \mathbf{P}(\theta)_{xt}\right]   &  =\left[
\begin{array}
[c]{cccc}%
\frac{1}{\cos^{2}\theta} & 0 & 0 & \pm\frac{\sin\theta}{\cos^{2}\theta}\\
0 & 1 & 0 & 0\\
0 & 0 & 1 & 0\\
\pm\frac{\sin\theta}{\cos^{2}\theta} & 0 & 0 & \frac{1}{\cos^{2}\theta}%
\end{array}
\right]
\end{align}

It is remarkable that the two universal Rayleigh-Taylor\_Kelvin-Helmholtz
fluid instability patterns discussed above are related to the matrix elements
of the non-linear Lorentz transformations. \ The two instabilities are
generated by the functions $Q=1/\cos^{2}(\theta)$ and $Q=\sin(\theta)/\cos
^{2}(\theta).$ \ These functions have another surprising property: \ they can
be put into correspondence with non-linear, conformal, Lorentz transformations
that generate rotations and expansions. \ 

For consider the angular coordinate being proportional to arclength to yield
the extended Lorentz transformation: The conformality factor is $\lambda
^{2}=1/\cos^{2}\theta$\ and the extended Lorentz transformation $\left[
\mathbb{P}\right]  $ is such that%

\begin{equation}
\left[  \mathbb{P}\right]  ^{T}\circ\left[  \eta_{\mu\nu}\right]  \circ\left[
\mathbb{P}\right]  \Rightarrow\lambda^{2}\cdot\left[  \eta_{\mu\nu}\right]  .
\end{equation}
The result implies that the zero set Eikonal equation is preserved invariantly
by the Poincare, or extended Lorentz transformations $\left[  P(\theta
)_{xt}\right]  $. \ In electromagnetic theory this means that the propagating
discontinuities remain invariant under conformal Lorentz transformations of
rotation and expansion, $\left[  \mathbb{P}\right]  $; \ 

\begin{remark}
The bottom line is that \textit{signals, defined as propagating
discontinuities \cite{Fock} \cite{Hadamard}, remain signals to all conformally
equivalent electromagnetic observers, not just Lorentz equivalent observers.}
\end{remark}

It is well known that the null quadratic form can also be related to Spinors,
treated as complex mappings of 4D to 4D. \ Such objects can be composed of
complex combinations of Hopf maps. \ The implication is that the conformal
Lorentz transformations, which are directly related to the instability
functions $Q=1/\cos^{2}(s)$ and $Q=\sin(s)/\cos^{2}(s)$\ of hydrodynamic
wakes, indicate that Spinors are not just artifacts of electromagnetic theory,
but also apply to hydrodynamic systems. \ This is a relatively unexplored area
of hydrodynamics and continuous topological evolution. \ The bottom line of
this result is that the relationship between the hydrodynamic instabilities
and non-linear Lorentz rotations can not be considered as an accident. \ Real
fluids have finite wave propagation velocities and are not perfectly
incompressible. \ These formal clues can be used to associate Spinors to wake
and instability formation, in terms of the known relationship between Spinors,
Harmonic vector fields and Minimal surfaces.

\subsection{The Cartan Connection matrix of 1-forms.}

Following the methods of section 2, above, the Cartan Connection is computed
for the Basis Frame of the extended Lorentz transformation, $\left[
\mathbf{P}(\theta(x,y,z,t))_{xt}\right]  .$\bigskip

\begin{center}%
\begin{align}
&  \text{\textbf{The Cartan Connection matrix }}\left[  \mathbb{C}\right]
\text{\textbf{\ of 1-forms based on}}\left[  \mathbb{P}(\theta)_{xt}\right] \\
\lbrack\mathbb{C}]_{\mathbf{P}(\theta)}  &  =\left[
\begin{array}
[c]{cccc}%
\sin(\theta)d\theta/\cos(\theta) & 0 & 0 & d\theta/\cos(\theta)\\
0 & \sin(\theta)d\theta/\cos(\theta) & 0 & 0\\
0 & 0 & \sin(\theta)d\theta/\cos(\theta) & 0\\
d\theta/\cos(\theta) & 0 & 0 & \sin(\theta)d\theta/\cos(\theta)
\end{array}
\right]
\end{align}

\end{center}

\subsection{The Vectors of Torsion 2-forms}

The rest of the presentation is extracted from Maple computations based upon
$\left[  \mathbb{P}(\theta)_{xt}\right]  .$ \ The function $\theta
=\theta(x,y,z,t)$ is unspecified:%

\begin{align}
&  \text{\textbf{The\ vector\ of\ 1-form\ Potentials }}\\
\left\vert A^{k}\right\rangle  &  =\left[  \mathbb{P}(\theta)_{xt}\right]
\circ\left\vert dx^{a}\right\rangle =\left\vert
\begin{array}
[c]{c}%
\{dx+\sin(\theta)dt\}/\cos(\theta)^{2}\\
dy/\cos(\theta)\\
dz/\cos(\theta)\\
\{\sin(\theta)dx+dt\}/\cos(\theta)^{2}%
\end{array}
\right\rangle
\end{align}

\begin{align}
&  \text{\textbf{The Vector of Intensity 2-forms \ }}\\
\text{\ }\left\vert F^{k}\right\rangle  &  =d\left\vert A^{k}\right\rangle
=\left\vert
\begin{array}
[c]{c}%
\{2d\theta\symbol{94}dx+(1+\sin(\theta)^{2})d\theta\symbol{94}dt\}\sin
(\theta)/\cos(\theta)^{3}\\
\sin(\theta)\{d\theta\symbol{94}dy\}/\cos(\theta)^{2}\\
\sin(\theta)\{d\theta\symbol{94}dz\}/\cos(\theta)^{2}\\
\{(1+\sin(\theta)^{2})d\theta\symbol{94}dx+2d\theta\symbol{94}dt\}\sin
(\theta)/\cos(\theta)^{3}%
\end{array}
\right\rangle
\end{align}

\begin{center}%
\begin{align}
&  \text{\textbf{The Vector of Excitation 2-forms \ \ (affine torsion)\ \ }}\\
\text{\ \ \ \ }\left\vert G^{a}\right\rangle  &  =\left[  \mathbb{C}\right]
\symbol{94}\left\vert dx^{m}\right\rangle =\left\vert
\begin{array}
[c]{c}%
\{\sin(\theta)d\theta\symbol{94}dx+d\theta\symbol{94}dt\}/\cos(\theta)\\
\sin(\theta)\{d\theta\symbol{94}dy\}/\cos(\theta)\\
\sin(\theta)\{d\theta\symbol{94}dz\}/\cos(\theta)\\
\{d\theta\symbol{94}dx+\sin(\theta)d\theta\symbol{94}dt\}/\cos(\theta)
\end{array}
\right\rangle
\end{align}

\begin{align}
\text{\textbf{The 4 Topological Torsion 3-forms }}  &  \mathbf{:}%
\text{(A\symbol{94}F)}^{k}\\
\text{Topological Torsion }\ \left\vert
\begin{array}
[c]{c}%
A^{1}\symbol{94}F^{1}\\
A^{2}\symbol{94}F^{2}\\
A^{3}\symbol{94}F^{3}\\
A^{4}\symbol{94}F^{4}%
\end{array}
\right\rangle  &  =\left\vert
\begin{array}
[c]{c}%
\{dx\symbol{94}d\theta\symbol{94}dt\}/\cos(\theta)^{3}\\
0\\
0\\
-\{dx\symbol{94}d\theta\symbol{94}dt\}/\cos(\theta)^{3}%
\end{array}
\right\rangle
\end{align}

\end{center}

It is remarkable that the individual components of Topological Torsion are not
zero, but the composite Total Topological Torsion vanishes. \ The Total second
Poincare Invariant vanishes.

\begin{center}%
\begin{align}
\text{\textbf{The 4 Topological Spin 3-forms }}  &  \mathbf{:}\text{{}%
}\mathbf{\ }(A\symbol{94}G)^{k}\\
\text{Topological Spin }\ \left\vert
\begin{array}
[c]{c}%
A^{1}\symbol{94}G^{1}\\
A^{2}\symbol{94}G^{2}\\
A^{3}\symbol{94}G^{3}\\
A^{4}\symbol{94}G^{4}%
\end{array}
\right\rangle  &  =\left\vert
\begin{array}
[c]{c}%
\{dx\symbol{94}d\theta\symbol{94}dt\}/\cos(\theta)^{3}\\
0\\
0\\
-\{dx\symbol{94}d\theta\symbol{94}dt\}/\cos(\theta)^{3}%
\end{array}
\right\rangle
\end{align}

\end{center}

Also, it is remarkable that the individual components of Topological Spin are
not zero, but the composite Total Topological Spin vanishes. \ The Total First
Poincare Invariant vanishes.

Note that the Congruent quadratic form, or metric, is always the Minkowski
metric, to within a factor. \ The Eikonal equation is an invariant of the
Lorentz system. \ With a constant conformal factor, the Riemann curvature of
the metric is zero. \ 

However, if the conformal factor is not zero, then the conformal metric can
exhibit curvature properties associated with expansions and rotations of space time.

The algebra of a Physical Vacuum in terms of a Basis Frame with the structural
format $\left[  \mathbb{P}(\theta)_{xt}\right]  $ is quite extensive. \ The
best place to see and study the results and derivations using the general
theory are to be found in the Maple programs:

\begin{center}
http://www22.pair.com/csdc/pdf/mapleEP3-Lorentzboost.pdf

http://www22.pair.com/csdc/pdf/mapleEP3-Lorentzwake.pdf
\end{center}

\section{Example 4. \ [$\mathbb{B}$] as a Projection in terms of the Hopf
map.}

The Hopf map is a famous mapping from 4D space to 3D space. \ One
representative of a Hopf mapping is given by the functions $\{X,Y,Z\}$and the
3 exact differentials $\{dX,dY,dZ\}$
\begin{align}
\left\vert
\begin{array}
[c]{c}%
X\\
Y\\
Z\\
\end{array}
\right\rangle  &  =\left\vert
\begin{array}
[c]{c}%
(xz+ys)\\
(xs-yz)\\
(x%
{{}^2}%
+y%
{{}^2}%
)/2-(z%
{{}^2}%
+s%
{{}^2}%
)/2\\
\end{array}
\right\rangle ,\\
\left\vert
\begin{array}
[c]{c}%
dX\\
dY\\
dZ\\
A_{Hopf}%
\end{array}
\right\rangle  &  =\left\vert
\begin{array}
[c]{c}%
zdx+xdz+sdy+yds\\
sdx+xds-zdy-ydz\\
xdx+ydy-zdz-sds\\
-xdy+ydx-zds+sdz
\end{array}
\right\rangle .
\end{align}
The additional 1-form
\begin{equation}
\text{Hopf 1-form \ \ }A_{Hopf}=-xdy+ydx-zds+sdz
\end{equation}
is added to the set of 3 perfect differentials to form a vector of 1-forms
that can be related to a Basis Frame of the type%
\begin{align}
\text{ }\left[  \mathbb{B}_{Hopf}\right]   &  =\left[
\begin{array}
[c]{cccc}%
z & s & x & y\\
s & -z & -y & x\\
x & y & -z & -s\\
-y & x & -s & z
\end{array}
\right]  ,\\
\left[  \mathbb{B}_{Hopf}\right]  \circ\left\vert
\begin{array}
[c]{c}%
dx\\
dy\\
dz\\
ds
\end{array}
\right\rangle  &  =\left\vert
\begin{array}
[c]{c}%
dX\\
dY\\
dZ\\
A_{Hopf}%
\end{array}
\right\rangle \\
\det\left[  \mathbb{B}_{Hopf}\right]   &  =+(x%
{{}^2}%
+y%
{{}^2}%
+z%
{{}^2}%
+s%
{{}^2}%
)^{2}=(R^{2})^{2}%
\end{align}
$\left[  \mathbb{B}_{Hopf}\right]  $ is in the format ready to be used for
infinitesimal mappings into nearby neighborhoods of a Physical Vacuum. \ \ The
Cartan Connection (based on $\left[  \mathbb{B}_{Hopf}\right]  ),$as a matrix
of 1-forms, becomes:

\begin{center}%
\begin{align}
&  \text{\textbf{Cartan Connection}}\\
\lbrack\mathbb{C}]_{\mathbf{P}(\theta)}  &  =\left[
\begin{array}
[c]{cccc}%
C_{1}^{1} & -C_{1}^{2} & -C_{1}^{3} & -C_{1}^{4}\\
C_{1}^{2} & C_{1}^{1} & -C_{1}^{4} & -C_{2}^{4}\\
C_{1}^{3} & C_{1}^{4} & C_{1}^{1} & -C_{1}^{2}\\
C_{1}^{4} & C_{2}^{4} & C_{1}^{2} & C_{1}^{1}%
\end{array}
\right] \\
C_{1}^{1}  &  =d\ln(R^{2})/2\\
C_{1}^{2}  &  =(sdz-zds+ydx-xdy)/R^{2}\\
C_{1}^{3}  &  =(sdy-yds~+xdz-zdx)/R^{2}\\
C_{1}^{4}  &  =(ydz-zdy~+xds-sdx)/R^{2}\\
C_{2}^{4}  &  =(yds-sdy~+zdx-xdz)/R^{2}%
\end{align}

\begin{align}
\text{\textbf{Hopf map}}  &  :\text{\textbf{The Congruent pullback metric }}\\
\left[  g_{jk}\right]  _{Hopf}  &  =[\mathbb{B}_{hopf\_transpose}]\circ
\eta\circ\lbrack\mathbb{B}_{hopf}],\\
\left[  g_{jk}\right]  _{Hopf}  &  =\left[
\begin{array}
[c]{cccc}%
g_{11} & -2xy & 2ys & -2yz\\
-2xy & g_{22} & -2xs & 2xz\\
2ys & -2xs & g_{33} & -2sz\\
-2yz & 2xz & -2sz & g_{44}%
\end{array}
\right]  ,\\
g_{11}  &  =(-x%
{{}^2}%
+y%
{{}^2}%
-z%
{{}^2}%
-s%
{{}^2}%
)\\
g_{22}  &  =(+x%
{{}^2}%
-y%
{{}^2}%
-z%
{{}^2}%
-s%
{{}^2}%
)\\
g_{33}  &  =(-x%
{{}^2}%
-y%
{{}^2}%
-z%
{{}^2}%
+s%
{{}^2}%
)\\
g_{44}  &  =(-x%
{{}^2}%
+y%
{{}^2}%
+z%
{{}^2}%
-s%
{{}^2}%
)
\end{align}

\begin{align}
\text{\textbf{Hopf Map }}  &  \mathbf{: }\text{Vector Array of \textbf{Affine
Torsion 2-forms }}\\
\text{Field Excitations }\left\vert
\begin{array}
[c]{c}%
G^{x}\\
G^{y}\\
G^{z}\\
G^{s}%
\end{array}
\right\rangle  &  =2/(R^{2})\left\vert
\begin{array}
[c]{c}%
ydy\symbol{94}dx+yds\symbol{94}dz\\
-xdy\symbol{94}dx-xds\symbol{94}dz\\
sdy\symbol{94}dx+sds\symbol{94}dz\\
-zdy\symbol{94}dx-zds\symbol{94}dz
\end{array}
\right\rangle ,~\text{(D and H)}\\
R^{2}  &  =x^{2}+y^{2}+z^{2}+s^{2}.
\end{align}

\begin{align}
\text{\textbf{Hopf Map }}  &  :\text{\textbf{Vector potentials} }\left\vert
A^{k}\right\rangle \\
\text{Potentials }\left\vert A^{k}\right\rangle  &  =\left\vert
\begin{array}
[c]{c}%
zdx+sdy+xdz+yds\\
sdx-zdy-ydz+xds\\
xdx+ydy-zdz-sds\\
zds-sdz+xdy-ydx
\end{array}
\right\rangle
\end{align}

\begin{align}
\text{\textbf{Hopf Map }}  &  :\text{\textbf{Intensities } }\left\vert
F^{k}\right\rangle =d\left\vert A^{k}\right\rangle \\
\text{Intensities}  &  \text{: \ }\left\vert F^{k}\right\rangle =\left\vert
\begin{array}
[c]{c}%
0\\
0\\
0\\
2dx\symbol{94}dy+2dz\symbol{94}ds
\end{array}
\right\rangle
\end{align}

\end{center}

Note that first three 2-forms of field intensities are zero, $F^{1}%
=F^{2}=F^{3}=0,$ but all four of the field excitations are not zero,
$G^{1}=G^{2}=G^{3}=G^{4}\neq0.$ \ It is also remarkable that only the last
Topological Torsion 3-form is not zero, but all 4 of the elements that make of
the Total Topological Spin 3-form are not zero. \ For these results and others
of more complicated algebraic display, see the Maple programs at

\begin{center}
http://www22.pair.com/csdc/pdf/mapleEP5-Hopf.pdf,

which will be published on a CD rom, \cite{vol6}.
\end{center}

\section{Thermodynamic Phase Functions from [$\mathbb{B}$]}

As pointed out in my first volume\footnote{See http://www.lulu.com/kiehn} "Non
Equilibrium Thermodynamics and Irreversible Processes", that, given any
matrix, useful information (related to Mean and Gaussian Curvatures of the
domain described by the matrix) can be obtained from the Cayley-Hamilton
Characteristic Polynomial. \ This polynomial exists for any square matrix.
\ In particular, for 4 x 4 matrices, if the matrix has no null eigenvalues
(maximal rank), the domain is then Symplectic and the characteristic
polynomial is of 4th degree in the polynomial variable (which represents the eigenvalues).

There exists a well known transformation of complex variable, $\xi=s+M/4,$
which will reformulate the characteristic polynomial into a system where the
"mean curvature" is zero:%

\begin{align}
\Theta(x,y,z,t;\ \xi)  &  =\xi^{4}-X_{M}\xi^{3}+Y_{G}\xi^{2}-Z_{A}\xi
^{1}+T_{K}\Rightarrow0,\\
\text{becomes \ \ \ \ }\Psi(x,y,z,t;\ s)  &  =s^{4}+gs^{2}-as+k=0.
\end{align}
\newline

Consider the reduced Phase formula, $\Psi=0$, and its derivatives with respect
to the family parameter, $s.$ \
\begin{align}
\Psi &  =s^{4}+gs^{2}-as+k=0,\\
&  \therefore k=-(s^{4}+gs^{2}-as),\\
\Psi_{s}  &  =\partial\Phi/\partial s=4s^{3}+2gs-a=0\\
&  \therefore a=4s^{3}+2gs\\
\Psi_{ss}  &  =\Phi_{s}=\partial^{2}\Phi/\partial s^{2}=12s^{2}+2g=0\\
&  \therefore g=-6s^{2}%
\end{align}

\subsection{The Higgs potential as an Envelope of the Thermodynamic Phase
function.}

Replacing the parameter $a$ (from\ the envelope condition, $\Psi_{s}=0)$ in
the equation for $k$ yields%
\begin{equation}
\text{\ }k=s^{2}(3s^{2}+g).
\end{equation}
A\ plot of this implicit surface appears below. \ The envelope of the reduced
quartic polynomial yields a thermodynamic phase function that establishes an
extraordinary correspondence between the binodal and spinodal lines of a van
der Waals gas, and its critical point. \ I call this envelope, the Higgs Phase
function. \ Of particular importance are the curves $s=0$ and $\partial
k/\partial s=0$ on the Higgs phase function. \ These curve $s=0$ defines the
thermodynamic Binodal line as a pitch-fork bifurcation set, and the set
$\partial k/\partial s=0$ defines the Spinodal line as the limit of phase
stability. \ The critical point is at $s=g=k=0.$ \ These topological concepts
can be mapped to, and should be compared with, the historic features of a van
der Waals gas.

\begin{center}%
{\includegraphics[
height=2.2952in,
width=3.039in
]%
{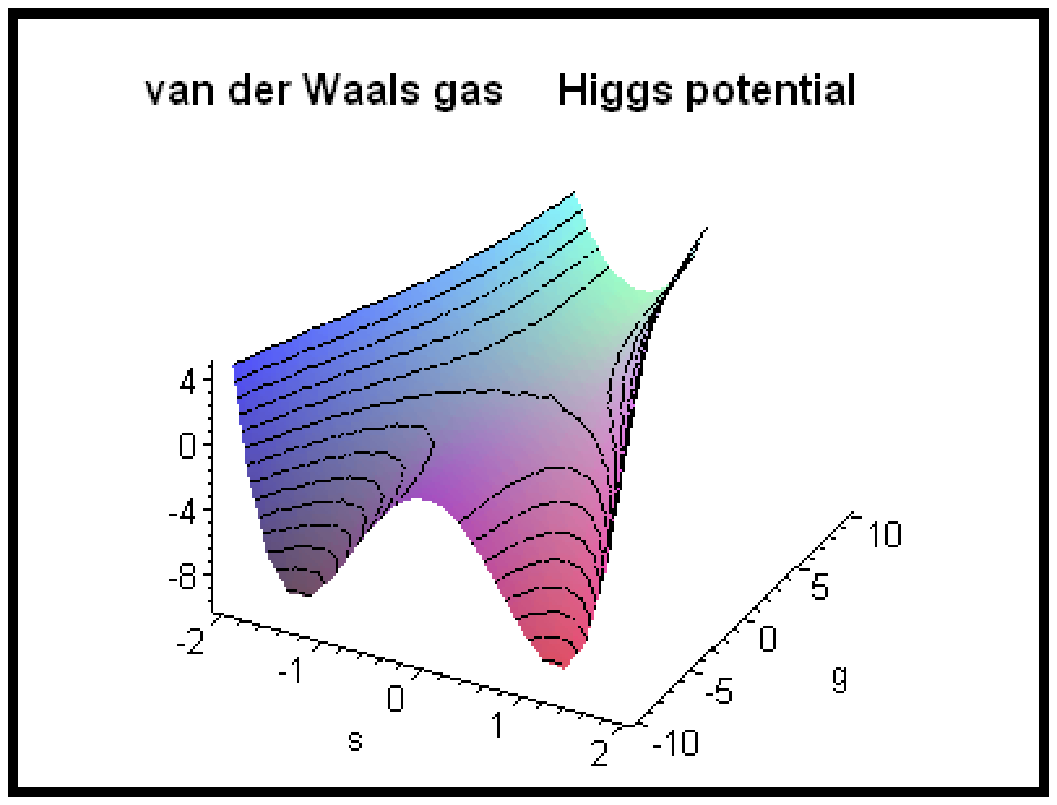}%
}%
\ 

\textbf{The vertical axis is the Higgs function, k.}

\textbf{Singularities in} $[\mathbb{B}]$ \textbf{occur when} $k=0$ \textbf{or}
$\partial k/\partial s=0.$

\textbf{The critical point is where} $k=s=g=0.$
\end{center}

The conjecture is that Higgs features (related to mass and inertia) have a
basis in topological thermodynamics, for at the critical point it is known
from chemical thermodynamics (and Lev Landau) that there are large
fluctuations in density, as the material attempts to condense, and these
fluctuations are correlated with a 1/r$^{2}$ force law. \ It should be noted
that the Higgs function, $k$, is related to the determinant of the Basis
Frame, and is an artifact of 4 topological dimensions. \ The $s$ component is
related to an abstract order parameter, or a molar (mass) density\footnote{The
effective eigen value, $s$, is related to the molar density minus the energy
of Mean curvature.} in a thermodynamic interpretation, and $g$ has the
abstract features of a temperature. \ Details are to be found in \cite{vol1}.
\ \ An interesting application is in the form of a cosmological model, where
it is presumed that the universe is a low density gas near its critical point
\cite{RMKcosmos}.

It is remarkable to me that all of this starts from the sole assumption of a
Basis Frame $\left[  \mathbb{B}\right]  .$ \ The Higgs idea and the extended
Yang Mills theory (and weak force) seem to be related to an irreducible Pfaff
topological dimension 4, where evolutionary irreversible processes are
possible and parity is not preserved. \ 

I conjecture that if the Phase function has 1 null eigen value, then the space
is related to a non equilibrium configuration of Pfaff dimension 3, for which
parity is always preserved (the strong Force). \ The electromagnetic domain
only requires Pfaff dimension 2, and the gravity domain is embedded in Pfaff
topological dimension 1. \ This follows from an old I argument (based on
differential geometry) that I presented in 1975 \cite{rmksubmersive}.

The idea is that spaces of Pfaff Topological dimension 2 or less (hence
thermodynamically these domains are isolated or are in thermodynamic
equilibrium) are topologically \textit{connected}, and interactions can range
over the entire domain (long range forces - gravity and electromagnetism).
\ However, spaces of Pfaff Dimension 3 or more are topologically
\textit{disconnected} domains of multiple components \cite{vol1}, and are not
in thermodynamic equilibrium. \ Hence interactions and forces are short range
(strong and weak forces, that may or may not preserve parity)

\end{document}